\documentclass[journal]{IEEEtran}

\ifCLASSINFOpdf
\else
\fi
\usepackage{cite}
\usepackage{amsmath,amssymb,amsfonts}
\usepackage{algorithmic}
\usepackage{graphicx}
\usepackage{textcomp}
\def\BibTeX{{\rm B\kern-.05em{\sc i\kern-.025em b}\kern-.08em
    T\kern-.1667em\lower.7ex\hbox{E}\kern-.125emX}}

\usepackage{balance}
\usepackage[bottom]{footmisc}
\usepackage{comment}

\usepackage{url}
\usepackage{kotex}
\usepackage{pifont}
\usepackage{multirow}
\usepackage{amssymb}
\usepackage{amsmath}
\usepackage{setspace}

\newcommand{\sgxssd}{\textit{SGX-SSD}}

\newcommand{\squishlist}{
	\begin{list}{$\bullet$}
		{ \setlength{\itemsep}{0pt}      \setlength{\parsep}{-0pt}
			\setlength{\topsep}{4pt}       \setlength{\partopsep}{0pt}
			\setlength{\listparindent}{-2pt}
			\setlength{\itemindent}{-5pt}
			\setlength{\leftmargin}{1em} \setlength{\labelwidth}{0em}
			\setlength{\labelsep}{0.5em} } }
	
	\newcommand{\squishend}{
\end{list}}

\newcommand{\myintertext}[1]{\end{enumerate} #1}
\newcommand{\myintertextend}{\begin{enumerate}[resume]}

\let\oldnl\nl
\newcommand{\nonl}{\renewcommand{\nl}{\let\nl\oldnl}}

\begin{document}

\title{A Policy-based Versioning SSD with Intel SGX}

\author{
	\IEEEauthorblockN{Jinwoo Ahn, Seungjin Lee, Jinhoon Lee, Youngwoo Ko, Junghee Lee, and Youngjae Kim}\\
    \thanks{J. Ahn, S. Lee, J. Lee, Y. Ko, and Y. Kim are with the Department of Computer Science and Engineering, Sogang University, Seoul, South Korea (e-mail: \{jinu37, seungjinn, kjjh1126, ppandol3, youkim\}@sogang.ac.kr).}
    \thanks{J. Lee is with the School of Cybersecurity, Korea University, Seoul, South Korea (e-mail: j\_lee@korea.ac.kr)}
	\thanks{Corresponding author: Youngjae Kim}
}
\maketitle

\begin{abstract}
Privileged malware neutralizes software-based versioning systems and destroys data.
To counter this threat, a versioning solid-state drive (SSD) that performs versioning inside the SSD has been studied. 
An SSD is a suitable candidate for data versioning because it can preserve previous versions without additional copying, and provide high security with a very small trusted computing base (TCB).
However, the versioning SSDs studied so far commonly use a full disk versioning method that preserves all file versions in a batch.
This paper demonstrates that SSDs, which provide full disk versioning, can be exposed to data tampering attacks when the retention time of data is less than the malware's dwell time.
To deal with this threat, we propose \sgxssd{}, a policy-based per-file versioning SSD to keep a deeper history for only the important files of users.
However, since the SSD isn't aware of a file semantic, and the versioning policy information should be securely received from the untrusted host computer, implementing the per-file versioning in SSD is a huge challenge.
To solve this problem, \sgxssd{} utilizes the Intel SGX and has a secure host interface to securely receive policy information (configuration values) from the user.
Also, to solve the file semantic unawareness problem of the SSD, a piggyback module is designed to give a file hint at the host layer, and an algorithm for selective versioning based on the policy is implemented in the SSD.
To prove our system, we prototyped \sgxssd{} the Jasmine OpenSSD platform in Linux environment. 
In the experimental evaluation, we proved that \sgxssd{} provides strong security with little additional overhead for selective per-file versioning.
\end{abstract}


\IEEEpeerreviewmaketitle

\section{Introduction}
\label{sec:intro}

To securely protect data from data corruption attacks by malware such as ransomware or wiper~\cite{wiper, wiper_petya, wiper_shammon_1}, one can use a file versioning software that records the file's history of changes~\cite{versioningFS, elephantfs, timetravelingdisk, ext3cow, traparray, backuptech, acronisbackup, snapshot}. 
Such file versioning software is implemented at the application or file system level.
In broad terms, there are two kinds of file versioning software, \textit{full disk versioning}~\cite{timetravelingdisk, bvssd, almanac} and \textit{per-file versioning}~\cite{versioningFS, elephantfs}. \textit{Full disk versioning} performs versioning on every file that is updated on storage. Through this, the whole disk can be recovered after a data corruption attack by malware. However, the space overhead of a storage can explode due to unnecessary versioning. On the other hand, \textit{per-file versioning} performs fine-grained versioning that only keeps the files selected by an user or application. This boosts the space efficiency of a disk since it blocks the backup of unnecessary files by setting a versioning policy for individual files. 
Moreover, important files can be versioned for a long time, and it reduces the security risks from malware data destruction attacks on a file.
 
However, the file versioning software mentioned above faces critical security challenges due to malware that is becoming more intelligent.
This intelligent malware, such as ring-0 privileged malware, uses the latest exploit techniques to escalate privilege~\cite{vulnerableFS} and compromise the system.
If malware acquires kernel privilege, anti-malware and file versioning services that are running on the host system for data protection will be neutralized. Moreover, to eliminate the possibility of data restoration, they neutralize the protection of file versioning software
by performing attacks aimed at the software backup device that is connected locally or remotely~\cite{ryuk2}. For example, Ryuk ransomware~\cite{ryuk2} exploits anti-recovery tools and shuts down the backup and anti-virus process.
In addition, the malware deletes not only the backup files of the infected computer, but also the remote backup files connected to the network.

To overcome this problem, recent studies have shifted their attention from the host side file-versioning software,
which can be compromised, to safely versioning the data inside a device~\cite{amoeba,SSDinsider,flashguard,ransomblocker,almanac, bvssd}. 
These studies commonly protect the integrity of data by retaining the past version of data inside the SSD\footnote{In this paper, this device is called a versioning SSD.}.
An SSD is a viable candidate to perform versioning for the following reasons~\cite{flashguard, almanac}. 
First, since SSD firmware is isolated from the host,
it is almost impossible to compromise the versioning system inside an SSD, even if malware escalates its privilege.
Second, when overwriting the data, an SSD doesn't erase the previous version of data, but rather logs it through an out-of-place mechanism~\cite{SSD}. This trait provides the chance to perform versioning without an additional copy overhead. 

The versioning SSDs~\cite{bvssd, almanac} being researched up to now perform \textit{full disk versioning}.
However, versioning all data inside the SSD with a limited space
incurs an huge space overhead. To alleviate this, the full disk versioning SSD implements an algorithm that actively 
deletes
the old version of data~\cite{bvssd, almanac}. 
However, 
the active deletion policy severely reduces the retention period of every file version. 
More importantly, a security loophole exists if the file's retention period is too short. 
When a file is tampered with or deleted by malware, the file can be permanently lost if the user does not notice it within the retention period~\cite{retention1, retention2}.
For example, suppose that intelligent ransomware evaded the detection and broke into the host system, and in the host system, a full disk versioning SSD that provides a retention period of three days is connected.
Ransomware can encrypt important files of the user, wait for three-day retention period to pass, and can ask for ransom on the fourth day. The user requests data restoration from the full versioning SSD, but it fails since the original data is already lost. 

In order to solve the aforementioned problems of full disk versioning SSDs, in this paper, we propose \sgxssd{}, a policy-based secure per-file versioning SSD. 
In SSD firmware, rather than applying the same policy to all files as in the full disk versioning SSD, the version policy is selectively applied to files based on the importance specified by the user or application. To the best of our knowledge, \sgxssd{} is the first attempt to provide \textit{per-file versioning} by the SSD firmware.
Specifically, \sgxssd{} implements a versioning system on a storage device rather than host-side software and protects data from potential threats from ring-0 privileged malware.
Also, it provides the file-based policy set by a user for selective per-file versioning on storage. Through a policy set, a user can configure the versioning policy of each file, such as the file retention period and backup cycle.
\sgxssd{} only manages the file version through the user-configured policy, preventing unnecessary backup of unimportant files and also providing strong security.

To implement a policy-based per-file versioning SSD, 
the following challenges should be solved.

\squishlist

\item
\textbf{Challenge\#1: Selective Per-file Versioning in SSD}
SSDs are not aware of the file semantic, but only accesses to data through logical block addresses. 
In addition, for \textit{per-file versioning}, the SSD must internally manage (create, delete, modify) versioning policies, and operate an algorithm to selectively determine whether to preserve each data page according to the policy.
So the question is, \textit{how can an user/application/OS tell the SSD that knows only the block semantics about the versioning policy for the data blocks in the file?} And the next follow-up question is \textit{how should SSD hardware be designed to internally perform per-file versioning?}

\item
\textbf{{Challenge\#2: Secure Host Interface on Compromised OS.}}
An SSD needs to securely get the versioning policy from the host.
If the host is compromised, the SSD can not trust the file management policy information that comes from the host. For example, privileged malware can mimic the authorized user and request policy deletion from the SSD to remove the past version of a file. The question is, \textit{how can an SSD safely get the policy-related input from the user even if the OS is compromised?}

\item
\textbf{{Challenge\#3: Performing Transparent and Autonomous File Versioning.}}
In the real world, multiple file I/Os are randomly performed on various applications. 
The SSD needs to manage versioning by identifying a file's policy whenever the file is updated.
The last question is, \textit{
how can it automate policy identification in the SSD while minimizing the user intervention?}

\squishend

\sgxssd{} adopts the following approaches to solve the challenges mentioned above.

\squishlist

\item
To overcome {{Challenge\#1}}, \sgxssd{} designs the \texttt{Piggyback Module}. 
When an application writes data to a file, file data is scattered to multiple logical blocks at the OS device driver and transferred to the SSD.
Whenever a user updates a file, the \texttt{Piggyback Module} appends the policy metadata to the end of the data buffer and sends it to the disk.
Therefore, even if an SSD doesn't know the file semantic, it can figure out the policy of an updated data page.
Also, the SSD implements a \texttt{Versioning Module} for \textit{per-file versioning}, 
which (i)internally manages policy metadata set by the host, and (ii) stores mapping information between each data page and policy, and (iii) runs an algorithm (\texttt{PV-Algorithm}) to determine whether to preserve each data page by referring to the policy during garbage collection. 
Details are described in Section~\ref{sec:pvssd}.

\item
To overcome {{Challenge\#2}}, \sgxssd{} builds a Secure Policy Manager (\texttt{SPM}) that provides a secure host interface to enable the policy management in an environment where an OS can be compromised.
With the \texttt{SPM}, only an authorized user can change or delete the policy.
The core of the \texttt{SPM} is to guarantee the secure transmission of the user's input from the host input device to the SSD. 
For this, \sgxssd{} implements the \texttt{SPM} inside a Trusted Execution Environment(TEE) provided by Intel SGX~\cite{costan2016intel}.
The \texttt{SPM} combines Aurora~\cite{aurora} with a two-way authentication module for secure communication with the SSD and ensures that only an authorized user can set policies.
\item

To overcome {{Challenge\#3}}, \sgxssd{} provides an autonomous file-to-policy mapping utility (\texttt{Policy Manager}).
A user configures an individual policy according to the type of a file (file extension, file name, directory, etc.). 
Once the policy is set, the \texttt{Policy Manager} automatically gives the proper policy to a file when the file is updated by an application. 
Also, the SSD gets the policy information and performs versioning according to that policy. User intervention and modifying the application are not required in this process. 

\squishend

We implemented \sgxssd{} on Linux environment using the Jasmine OpenSSD platform~\cite{jasmine}.
In order to analyze the performance overhead of \sgxssd{}, we evaluated the overhead of passing per-file information in the kernel to the SSD and the secure interface overhead. In addition, we measured the I/O performance of \sgxssd{} using a combination of synthetic and realistic workloads.
From the evaluation, we proved that \sgxssd{} shows negligible additional overhead for selective per-file versioning compared to a system without versioning, while offering strong security. 
\section{Background and Motivation}
\label{sec:back}

\subsection{Intel SGX}

Intel SGX~\cite{costan2016intel} is an instruction set provided by an Intel processor of Skylake or higher. 
Through the Trusted Execution Environment (TEE) called Enclave, SGX ensures the confidentiality and integrity of applications even when the OS is compromised.
The software developer divides the application into untrusted parts and Enclave parts, and implements an interface between the two parts.
The untrusted part calls the Enclave through ECALL, and the Enclave calls the untrusted part through OCALL.
For using hardware resources, since the Enclave cannot directly call system calls, it should execute OCALL first and request the untrusted part to call the system call.
Also, Intel SGX cannot perform safe I/O with external devices.
For example, values entered by the user through a user interface such as a keyboard are passed through the driver of the untrusted OS to reach the Enclave.
Accordingly, when the OS is compromised, there is a limitation that SGX cannot trust user inputs through the user interface (UI).

To overcome this, Aurora~\cite{aurora} uses System Management Mode (SMM), another privileged mode provided by the processor, to ensure secure I/O between the Enclave and external devices.
Aurora fundamentally blocks the possibility of man-in-the-middle attacks by privileged malware by excluding the kernel from the data transmission procedure between an external device and Enclave. 
Thus, Aurora guarantees that the user's input is safely transmitted to Enclave through the UI. 
Aurora safely gets the user's input through the following flow.
Aurora is composed of Enclave, which is an arrival part of the user's input, SMVisor that runs on SMM, and the secure session between the SMVisor and Enclave.
First, before receiving the user's input through a keyboard, Enclave invokes System Management Interrupt (SMI) through an OCALL, context switches to SMM, and SMVisor starts running.
In SMM mode, SMVisor reroutes the keyboard IRQ register value of an I/O APIC to the SMI handler.
Thus, when a keyboard input comes in afterwards, context is switched to SMM and the input value comes to SMVisor.
When SMVisor gets the user's input, it securely transmits it through the secure session established in advance.

\begin{figure}[!t]
	\begin{center}
		\begin{tabular}{@{}c@{}c@{}c@{}c@{}}
			\includegraphics[width=0.35\textwidth]{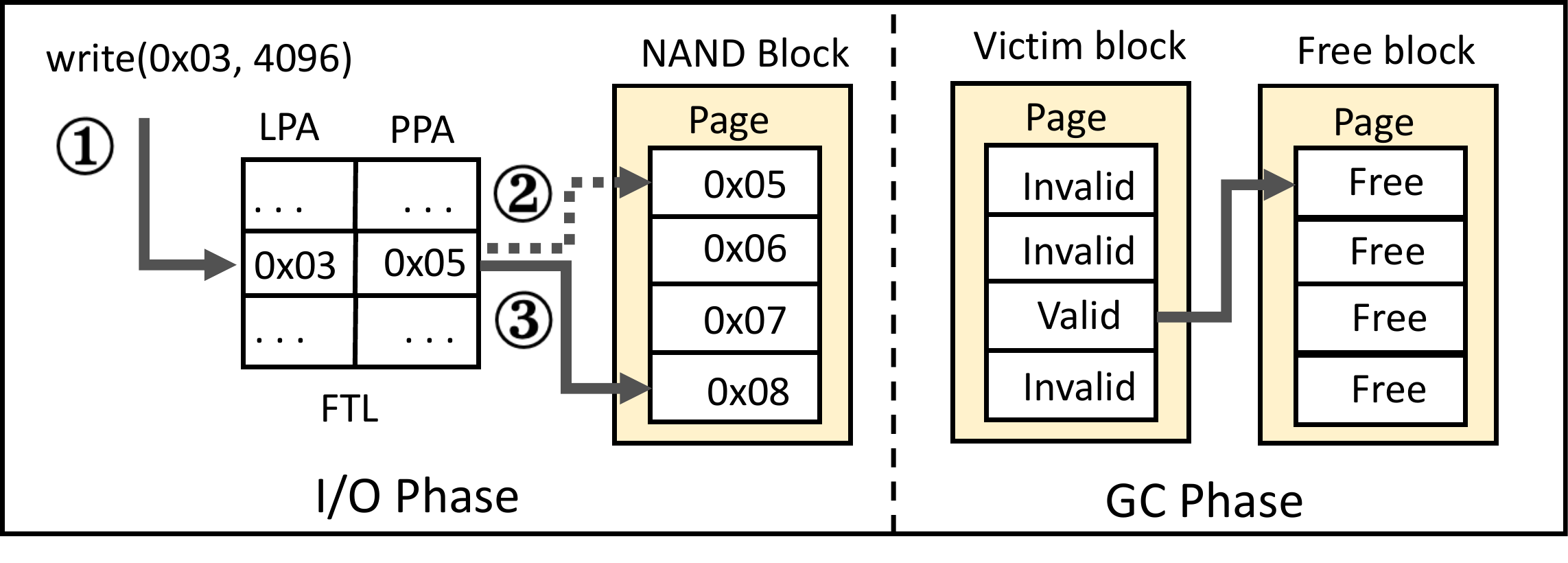} 
		\end{tabular}
		\caption{\small  I/O phase (out-of-place update) and GC phase in SSD.}	
		\label{fig:ssd}
	\end{center}
\end{figure}

\subsection{Solid-State Drive}

A NAND flash memory-based SSD is a storage storage device that replaces HDDs. An SSD provides higher I/O throughput and lower latency than an HDD, but has some limitations as follows.
First, while the SSD internally reads and writes NAND flash memory in units of pages, it can erase only in units of blocks composed of several pages. Block erase operations are hundreds to thousands of times slower than page read/write operations.
Second, pages that have already been written cannot be overwritten in place.
In other words, to overwrite a page, the page must be freed through a block erase operation.
However, it is very inefficient to erase the entire block containing the page to overwrite the page each time.
To solve this problem, the SSD controller internally performs out-of-place update operations and garbage collection (GC).

\textbf{Out-of-Place Update:}
Instead of writing data directly to the page address requested by the host, the SSD performs out-of-place updates through indirection.
Figure~\ref{fig:ssd} shows the workflow of an out-of-place update and GC operations.
The SSD runs a flash translation layer (FTL) in its internal memory, a table that maps logical page addresses (LPA) and physical page addresses (PPA) known to the host.
For Figure~\ref{fig:ssd}, 
suppose that LPA 0x03 is mapped to PPA 0x05 in the FTL, and the host requests to write 4 KB of data to LPA 0x03 (\ding{172}).
In the physical page of PPA address 0x05, valid data corresponding to LPA address 0x03 is stored. (\ding{173}).
Instead of erasing the page at PPA 0x05 using an expensive block erase operation, the SSD marks the physical page as invalid.
Then, the SSD is allocated a free physical page, writes the requested data to the page, and updates the corresponding entry in the FTL (\ding{174}). 

\textbf{Garbage collection:}
GC refers to the operation of producing free pages by deleting physical pages marked as invalid.
Generally, the GC phase is executed when the number of invalid pages in the SSD exceeds a certain threshold. 
In the GC phase, the SSD controller selects the victim block with the largest number of invalid pages among the NAND blocks.
In Figure~\ref{fig:ssd}, the selected victim block has three invalid pages and one valid page.
When performing GC, valid pages in the victim block are copied to the pre-allocated free block.
When copying is complete, the entire victim block is erased and a new free block is created.

\subsection{Versioning SSD}

The versioning SSD~\cite{flashguard, SSDinsider, amoeba, ransomblocker, almanac} implements a versioning system that retains the previous version of data inside the SSD firmware for a certain amount of time.
Additional storage devices are not required since data versioning is done inside the SSD.
The versioning SSD is implemented using the following characteristics of the SSD.
First, when overwriting the page, the SSD offers an out-of-place update operation that allocates a new page to write, not overwriting the original one~\cite{SSD}.
When an overwrite request occurs, the page in the SSD where the past data is stored becomes invalid, and is later erased during the GC phase.
By utilizing the out-of-place update operations of the SSD, the versioning SSD considers and preserves an invalid page as Old Version (\texttt{OV}) for recovery.
In this way, \texttt{OV} is preserved without additional copy overhead in the SSD.
Second, SSD firmware is stronger in terms of security than the upper-level host system for the following reasons~\cite{almanac, flashguard}.
(i) Since the firmware code of the SSD is isolated from the upper-level system, such as an OS, it is safe even if the OS is compromised.
(ii) The isolated firmware region has a much smaller trusted computing base (TCB) than the upper-level system.
(iii) The SSD is the place where data is stored, and it is the last barrier to protect the data.

The implementation of the versioning SSD is divided into a full disk versioning SSD~\cite{bvssd, almanac}, which backs up all data that is updated to the SSD for a certain period of time, and an SSD that performs ransomware detection and backup of selected blocks together~\cite{amoeba,SSDinsider,flashguard,ransomblocker}.
BVSSD~\cite{bvssd} is an early full disk versioning SSD research that utilizes the fact that the SSD preserves the data for a certain amount of time (about two days~\cite{bvssd}) before the time when garbage collection happens. 
BVSSD~\cite{bvssd} restores the previous version of data by logging the metadata alteration of the FTL. 
However, from the time after the garbage collection happens, which deletes the data, data restoration is not possible.
On the contrary, Project Almanac~\cite{almanac}, which is a state-of-the-art versioning SSD, maintains performance and retains data for a longer time than BVSSD by dynamically modifying the retention period.
Project Almanac basically monitors garbage collection overhead while preserving all versions of data.
When garbage collection overhead grows due to scarce space that results from \texttt{OV} preservation, space is made by erasing the oldest version of data. 
Thus, Project Almanac is a system that dynamically controls the retention period of each data by a workload's I/O intensity to secure the space in the SSD (minimum 3 days - maximum 56 days).

On the other hand, SSDs that perform ransomware detection and versioning together~\cite{amoeba,SSDinsider,flashguard,ransomblocker} are equipped with a ransomware detection module inside the device to back up only blocks with high risk of attack.
The ransomware detector determines the infection of data based on the encryption patterns~\cite{cryptolock} that ransomware have in common.
When an infection is detected, I/O stops and the data restoration process runs.
However, an intelligent malware can cleverly bypass the ransomware detector by corrupting the data with an unknown pattern. Also, defending against a wiper attack \cite{wiper, wiper_petya, wiper_shammon_1}, which doesn't show a specific encryption pattern and aims to indiscriminately destroy the data, is almost impossible.

All the previous versioning SSD research performed the block-level backup and did not perform the per-file versioning selected by a user.
Therefore, in this paper, we propose a per-file versioning SSD that can preserve files based on a user-configured policy.

\subsection{Motivation}
\label{sec:motivation}

Full disk versioning SSDs back up every version of data and incur an huge space overhead.
To ease this, an algorithm that dynamically limits the retention period of data is implemented in the full disk versioning SSD.
However, the limited retention period of a full disk versioning SSD leads to a security loophole.
To explain this problem, we define the terms pertaining to versioning in Table~\ref{tab:Declare}.
If malware with a dwell time(\texttt{DT}) longer than the full disk versioning SSD's minimum retention time(\texttt{RT}) breaks into the host system, the full disk versioning SSD can not guarantee the integrity of the whole storage data. This is because, when the user detects malware and requests file restoration, the retention period of original file data has already passed.

\begin{table}[!t]
	\centering	
	\small
	\caption{\small Definition of Terms for Versioning Systems.}
	\vspace{-0.2in}
	\begin{center}
		\resizebox{\columnwidth}{!}{
			\begin{tabular}{|l||p{7cm}|c|c|c|c|c|}
				\hline
				\textbf{Term}&\textbf{Description}\\
				\hline
				\hline
				Retention Time ({\texttt{RT}})&Period to guarantee recovery after data becomes invalid\\
			    \hline
			    \multirow{2}{*}{Dwell Time ({\texttt{DT}})}&The period of time between malware executing within an environment and it being detected\\
			    \hline
			    Old Version ({\texttt{OV}})&Previous version of data that is preserved for recovery\\
			    \hline
		\end{tabular}}
	    \label{tab:Declare}
	\end{center} 
\end{table}

\begin{figure}[!t]
	\begin{center}
		\begin{tabular}{@{}c@{}c@{}c@{}c@{}}
			\includegraphics[width=0.35\textwidth]{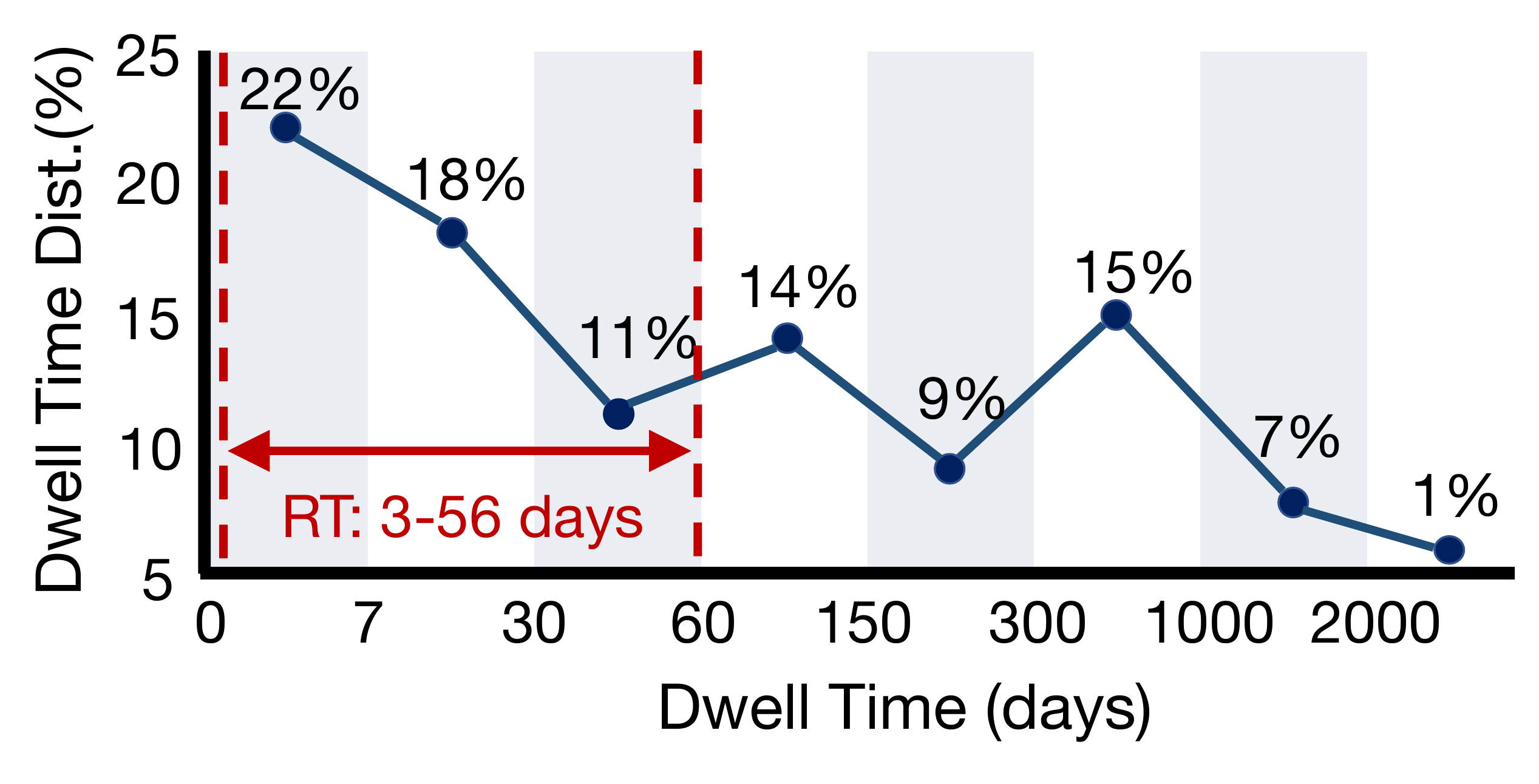} 
		\end{tabular}
		\caption{\small Global Dwell Time Distribution of Malware~\cite{dwelltime}.}	
		\label{fig:dwelltime}
	\end{center}
\end{figure}

\textit{The essence of this problem is that the \texttt{DT} of most malware is far longer than the \texttt{RT} of the full disk versioning SSD.}
Figure~\ref{fig:dwelltime} shows the statistical distribution of malware (\texttt{DT}). 
As shown in the figure, more than 50\% of attacks have a \texttt{DT} of 60 days or longer.
Also, more than 80\% of malware has a \texttt{DT} of 3 days of more.
On the other hand, Project Almanac~\cite{almanac}'s RT ranges from a minimum of 3 days to a maximum of 56 days.
This indicates that Project Almanac, which provides the longest \texttt{RT} among the full disk versioning SSD can be vulnerable to 50\% to 80\% of malware.

\begin{figure}[!t]
	\centering
    
    \begin{tabular}{c}
        \includegraphics[width=0.37\textwidth]{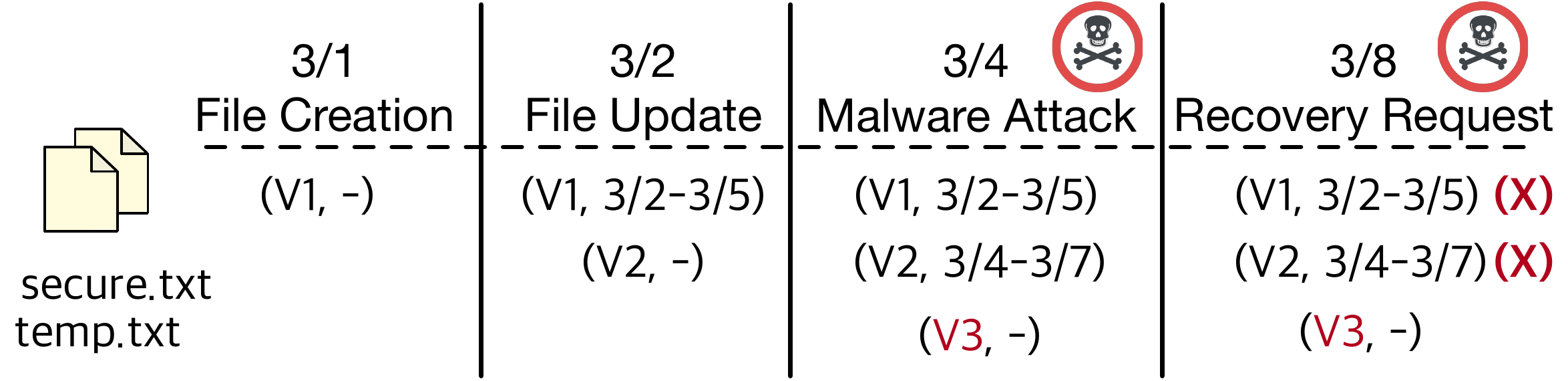}\\
        \small (a) Delayed Attack in Project Almanac\\
            \includegraphics[width=0.37\textwidth]{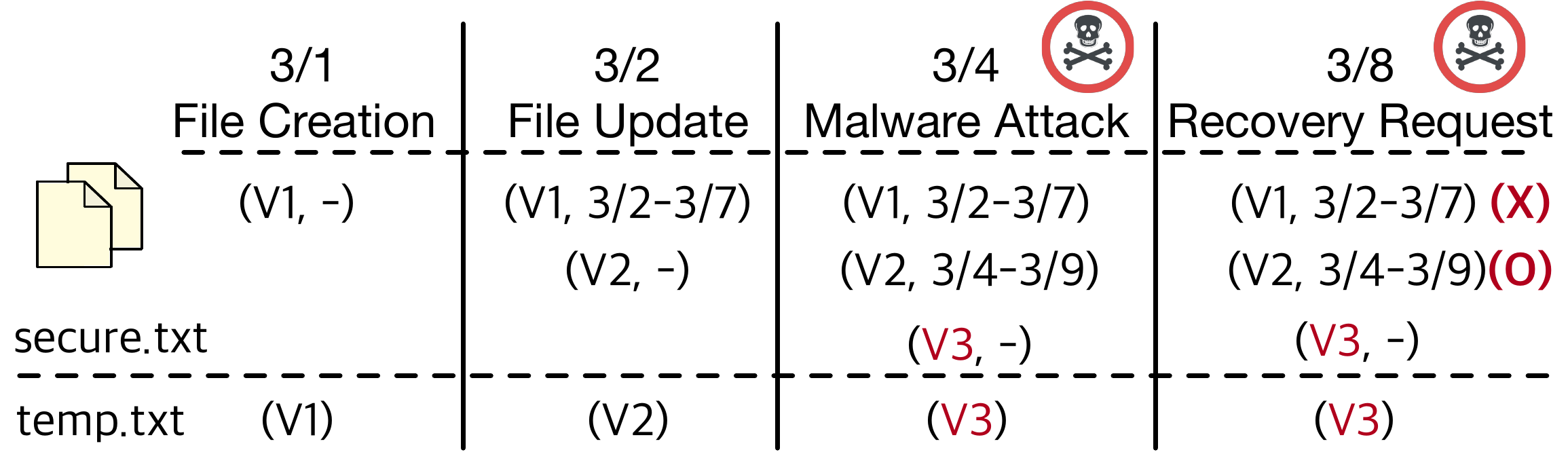}\\
        \small (b) Delayed Attack in \sgxssd{}
        \\
        \end{tabular}
      \caption{\small 
      Examples of Delayed Attack Scenarios.
      }
	\label{fig:delayed_attack}
\end{figure}

\textbf{Examples:}
We describes concrete examples for the aforementioned issues in Figure~\ref{fig:delayed_attack}.
Figure~\ref{fig:delayed_attack}(a) shows a scenario in which malware performs a delayed attack in Project Almanac.
Suppose {\texttt{secure.txt}} is an important file to be protected by the user, and {\texttt{temp.txt}} is a temporary file that does not need to be protected.
Malware increases I/O intensity to the SSD by writing dummy files,
so that Project Almanac's \texttt{RT} 
reaches the minimum \texttt{RT} (3 days).
Assume that the malware's {\texttt{DT}} is 4 days.
The user creates files ({\texttt{secure.txt}} and {\texttt{temp.txt}}) on 3/1 (March 1) in the Project Almanac environment and updates the files on 3/2.
At this point, the file data (V1) generated on 3/1 becomes the {\texttt{OV}}, and the newly updated data (V2) becomes valid data.
On 3/4, malware with a {\texttt{DT}} of 4 days invades the host system and encrypts the files ({\texttt{secure.txt}} and {\texttt{temp.txt}}).
At this time, there are two {\texttt{OV}} (V1, V2) on the SSD, and the data (V3) encrypted by malware becomes valid data.
The {\texttt{RT}} of V1 ends on 3/5, and the {\texttt{RT}} of V2 ends on 3/7, and V1 and V2 are erased.
The user detects malware on 3/8 and tries to recover data.
However, since the data (V1, V2) stored before the malware invasion does not remain on the SSD, both \texttt{secure.txt} and \texttt{temp.txt} fail to be recovered.

Figure ~\ref{fig:delayed_attack}(b) describes a scenario where malware performs a delayed attack in \sgxssd{}.
It is assumed that the user sets the {\texttt{RT}} of the important {\texttt{secure.txt}} to 5 days,
and {\texttt{RT}} is not set in {\texttt{temp.txt}} because its importance is low.
The user updates the data with the flow described above, and the malware attempts to encrypt it in the same way.
On 3/8 when malware is detected, the user requests data recovery.
{\texttt{temp.txt}} cannot be recovered because there is no \texttt{OV}.
On the other hand, \texttt{secure.txt} has an \texttt{RT} of 5 days, and V2 has not yet expired.
Therefore, the user can recover \texttt{secure.txt} at the point in time before the malware entered.
In addition, on 3/8, Project Almanac has two versions of \texttt{secure.txt} and \texttt{temp.txt}, so there is a total of four \texttt{OV}s.
On the other hand, \sgxssd{} does not have the \texttt{OV} of \texttt{temp.txt}, so it only have 2 \texttt{OV}s. 
This shows that the space resources obtained by reducing the \texttt{RT} of the less important file (\texttt{temp.txt}) can be used to protect the important file (\texttt{secure.txt}) for a longer period.

In summary, if the per-file versioning is possible in the SSD, different retention periods can be set for different files to balance data retention period and storage performance without adjusting the dynamic retention period.
Therefore, prolonged preservation of an important file is possible by reducing the unnecessary backup of less important files. 

\section{System Overview of \sgxssd{}}
\label{sec:overview}

\subsection{Threat Model}
\label{sec:threat}

We assume that malware can acquire the highest privilege (ring-0 level) and extract secrets from or make changes to the kernel, application software and data in the memory, unless they are explicitly protected by SGX.
Malware aims to delete or tamper with sensitive data of the user.
However, the hardware and its manufacturer are trusted.
This is a typical threat model of the trusted hardware.
We exclude attacks through physical access from our threat model.

\begin{figure}[!t]
	\begin{center}
		\begin{tabular}{@{}c@{}c@{}c@{}c@{}}
			\includegraphics[width=0.28\textwidth]{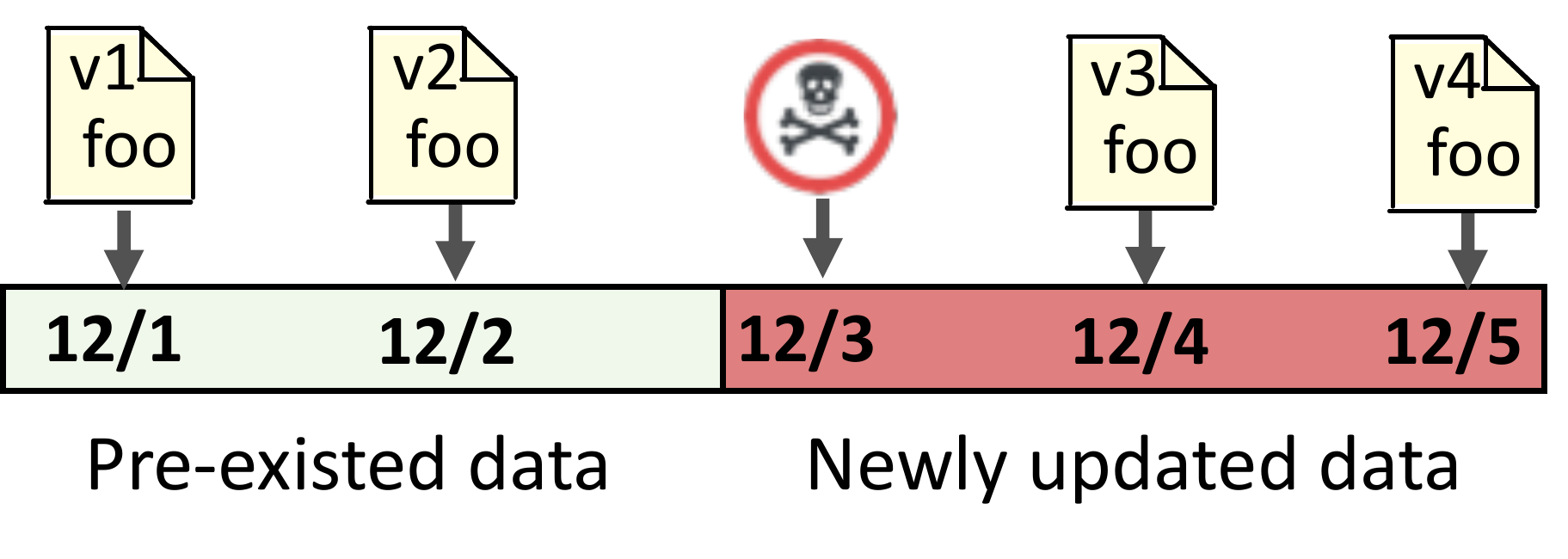} 
		\end{tabular}
		\caption{\small
		Description of the range of file versions that \sgxssd{} guarantees recovery}
		\label{fig:threatmodel}
	\end{center}
\end{figure}

To counter the threat of malware, \sgxssd{} guarantees the integrity of the all old versions of data stored on a disk based on the policy.
However, \sgxssd{} does not guarantee the integrity of newly updated data after malware infection. 
What we guarantee is limited to pre-existing data that was updated to the disk before malware infection.
This way, when the data is tampered with, the user recovers the tampered files up to the point in time before the malware invades.

Figure~\ref{fig:threatmodel} describes the backup version of a file guaranteed by the aforementioned \sgxssd{}.
Since the malware invaded on 12/3 (December 3), all file versions (v3, v4) newly updated by users can be compromised by malware.
On the other hand, \sgxssd{} guarantees data integrity of the version (v1, v2) prior to the malware invasion, so files can be restored with v1 or v2.
This is not a limitation of our system; but various versioning systems and backup solutions~\cite{versioningFS, elephantfs, timetravelingdisk, ext3cow, traparray, backuptech, acronisbackup, snapshot} provide the same level of security.
Thus, according to the policy, the integrity of files that have already expired is not guaranteed.
Since the user recognizes this, one can allocate a policy with a sufficiently long \texttt{RT} to protect data if the file is important.

SGX and SSD both have very small TCBs for TEE, so they are assumed to be malware-free.
Also, the user trusts the vendor that distributes \sgxssd{}, and assumes that the user's system is not infected with malicious code when the user first installs \sgxssd{}.
The goal of \sgxssd{} is to guarantee the integrity of file data, and it can not defend against side-channel attacks and DoS attacks. 
For example, a malware attack that depletes the storage space by dummy file write is possible, but it can not harm the \texttt{OV} of a file.
In \sgxssd{}, the capacity of the SSD can fill up with many {\texttt{OV}s} of a file. 
When SSD is full, \sgxssd{} stops I/O to protect the \texttt{OV}.
At that point, the user should delete the \texttt{OV} of a file and make space.

\subsection{Versioning policy and configuration}
\label{sec:retention_policy}

\sgxssd{} provides APIs or utilities that allow users or applications to directly set version policy configurations (version policy creation, deletion, and modification) on the files to directly manage file backup.
When creating a new versioning policy, a user specifies (i) configuration parameters and (ii) file rules.
A configuration parameter is a version control option for how the user will version the file.
On the other hand, the file version rule is a rule for deciding which file should be versioned.

\subsubsection{Configuration parameter}

The per-file policy provided by \sgxssd{} can specify the versioning strategy according to the following versioning options.

\squishlist
\item
{Retention Time (\texttt{RT}):}
A user configures the retention period of a file.
\sgxssd{} performs versioning of every update in real time during the retention period of a file.
Thus, within the file retention period, which is configured by a user, restoring the \texttt{OV} of a file is guaranteed.

\item
{Backup Cycle (\texttt{BC}):}
A user can determine the backup cycle of a file.
For example, if the backup cycle is set to one day, \sgxssd{} retains only one \texttt{OV} per day.
By using \texttt{BC}, the space overhead for preserving the \texttt{OV} of the SSD can be greatly reduced.

\squishend

\subsubsection{File Rule}
Users can use regular expressions to set rules for files corresponding to each policy.
Accordingly, a user can set a policy on a file/directory according to the file name, file extension, and directory. 

\squishlist
\item
\textit{Policy configuration based on file name:}
Users can impose versioning policies based on the file name.
Companies can effectively version various digital assets (contracts, design data, source code, important document files) or compliance (medical records, personal information) in accordance with their versioning policy.
In addition, database files, system files, and website files for service can also be individually versioned in real time according to policies.
\item

\textit{Policy configuration based on file extension:}
Many ransomware classify and encrypts users' important files according to their extensions~\cite{ransomaretarget}.
To defend against this, the user sets a safe version policy to keep the deep history of important files that ransomware frequently attacks.
A user can register a policy and set files of extension such as .doc, .xls, .pdf, .jpg, and .zip, which ransomware usually attacks, as a file rule and set the \texttt{RT} to one year as a policy configuration parameter.
For example, a user can create a policy rule and request versioning of *.pdf files as follows.

\begin{equation}
\label{eq:v2}
\footnotesize
\hspace{-0.2in} \$\{PolicyCreate\} \{FileRule=*.pdf\} \{RT=1year\} \{BC=1day\}
\end{equation}

From the time the policy is created,  
\sgxssd{} will guarantee a retention period of one year for the files that are mapped to the \texttt{*.pdf} regular expression.
However, since the backup cycle is one day, versioning is not performed every time a file is updated, and one version of the file is preserved per day.

\item
\textit{Policy configuration based on directory:}
To preserve a file for a long time, a user can select a secure directory.
A user sets the file rule by including the specific directory name, and sets the configuration parameter of the policy to have a sufficiently long retention period.
For example, a safe versioning policy can be set to the desktop directory that the user usually uses, or the directory that holds an important system file. 
Afterward, when the files saved in that directory are updated, automatic versioning is performed according to the registered policy.
\squishend

\subsection{System Software Architecture}
\label{sec:impl}

\begin{figure}[!t]
	\begin{center}
		\begin{tabular}{@{}c@{}c@{}c@{}c@{}}
			\includegraphics[width=0.4\textwidth]{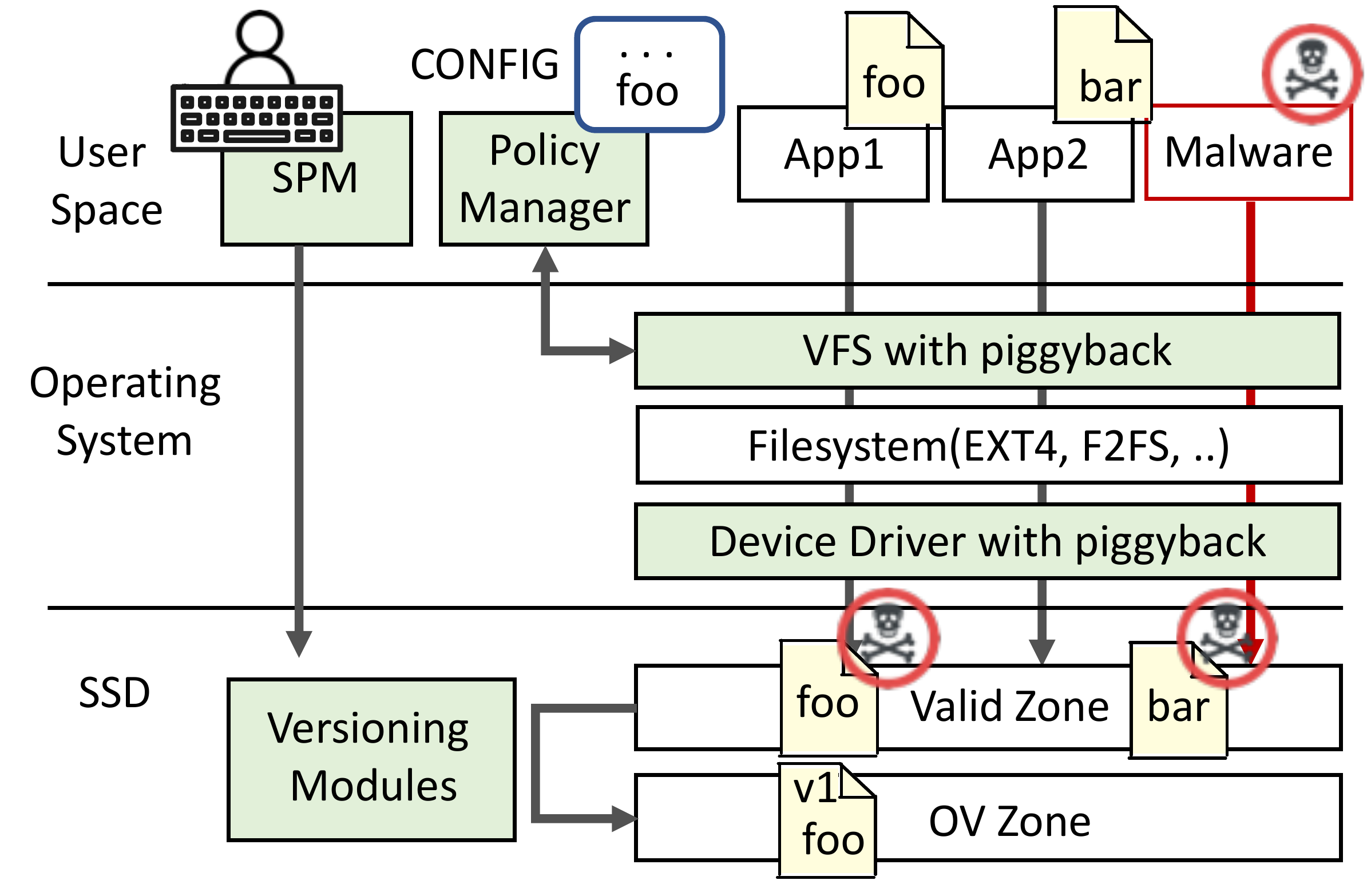} 
		\end{tabular}
		\caption{\small An Architecture of \sgxssd{}. \texttt{foo} is a secure file versioned by the policy, \texttt{bar} is a normal file that is not versioned since the policy doesn't exist.}
		
			\label{fig:architecture}
	\end{center}
\end{figure}

Figure~\ref{fig:architecture} shows an overview of the \sgxssd{} system software architecture.
\sgxssd{} is composed of  
(i) a secure policy delivery framework running at the host level and (ii) an \texttt{SSD Versioning Module} running inside the SSD.
The secure policy delivery framework allows the user to securely set the versioning policy in \sgxssd{} and delivers policy hints of blocks mapped to files updated by the application to the SSD. 
The secure policy delivery framework consists of three modules:
(i) \texttt{SPM}, (ii) \texttt{Policy Manager}, and (iii) \texttt{Piggyback Module}

\squishlist
\item
\texttt{SPM} is a safe utility based on SGX for a user's policy management, and it is responsible for creating, modifying, and deleting the versioning policy registered in the SSD.
\item
\texttt{Policy Manager} is the utility that saves the file rule mapped to the policy.
Based on the file rule, the \texttt{Policy Manager} reads the policy information corresponding to the file and sends it to the \texttt{Piggyback Module}.
\item
\texttt{Piggyback Module} is an OS kernel module.
When the application updates the file, the policy information provided by the \texttt{Policy Manager} is piggybacked to the data page of the updated file and transferred to the SSD.
\squishend

Apart from the above secure policy delivery framework, \texttt{SSD Versioning Module} is a software module implemented in SSD firmware that performs versioning on the files by the versioning policy.
The SSD is partition into two regions: 
$Valid Zone$ and $OV Zone$. $Valid Zone$ is a place where user-updated valid data is saved, while $OV Zone$ only saves the data that is in the $OV$ state.
Since the $OV Zone$ is not directly mapped to the LPA, access from the host is impossible.
Thus, even if malware compromises the host OS, it cannot tamper with the file's \texttt{OV}.
For example, unlike a normal file \texttt{bar} that is not versioned, even if malware overwrites the versioned secure file \texttt{foo} with encrypted data, the original version of the data (v1) is securely stored in the \texttt{OV Zone}. 

\subsection{\sgxssd{} Software Installation}
When deploying \sgxssd{}, the vendor hardcodes and hides a unique device key ($K_{dev}$) in \texttt{SPM} and SSD firmware. 
The vendor provides users with a security policy delivery framework, which contains an SGX-based safe utility (\texttt{SPM}), \texttt{Policy Manager}, and \texttt{Piggyback Module}. 
Also, the vendor provides the versioning SSD. 
Vendors may sell SSDs implemented with an \texttt{SSD Versioning Module} in hardware, but firmware updates of normal SSDs may allow users to securely download the \texttt{SSD Versioning Module} to their SSD devices~\cite{shade2011implementing}.
When the user installs \sgxssd{}, the user configures the size of backup space to store \texttt{OV}.
For example, if the user allocates 100 GB as a backup space in SSD which has 500 GB capacity, the user can use 400 GB for storing the valid data, and is able to save 100 GB for storing \texttt{OV}. 
If data in the backup space is compressed, more than 100 GB of data can be stored in the backup space. 
If the total size of \texttt{OV} exceeds the backup area, SSD stops I/O to preserve \texttt{OV}.
\section{Safe Policy Delivery Framework}

\subsection{Secure Policy Management}
\begin{figure}[!t]
	\begin{center}
		\begin{tabular}{@{}c@{}c@{}c@{}c@{}}
			\includegraphics[width=0.4\textwidth]{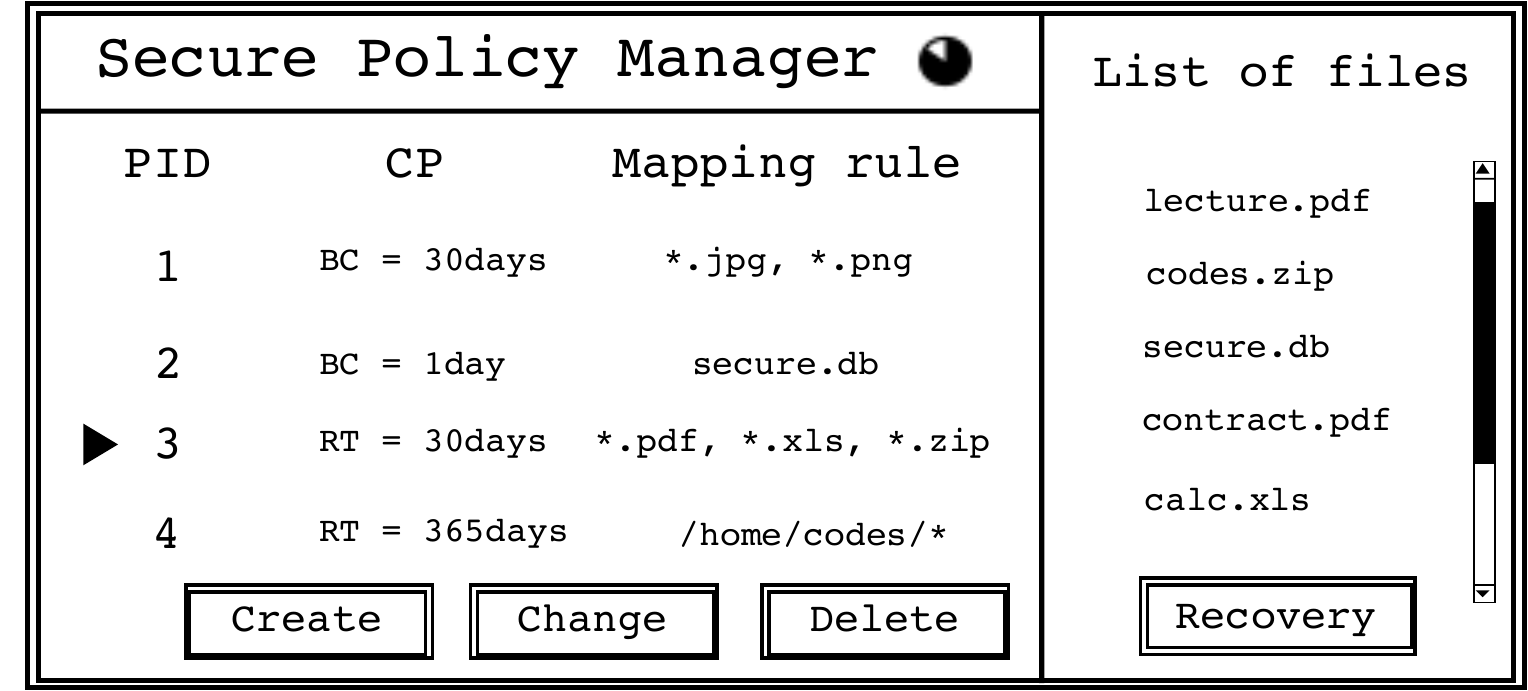} 
		\end{tabular}
 		\vspace{-0.05in}
		\caption{\small A secure user screen of the 
		\texttt{SPM} for policy configuration.
			}	\label{fig:spm}
 		\vspace{-0.15in}
	\end{center}
\end{figure}

When a user runs the \texttt{SPM}, it displays the information of the currently registered policy metadata and the lists of files versioned by the policy to the user, as in Figure~\ref{fig:spm}.
Each policy metadata managed by the \texttt{SPM} contains a policy identifier($Id_{(P_i)}$), configuration parameter($CP_{(P_i)}$) for the retention period (RT), backup cycle, and rules of files to be versioned by policy ($Rule_{(P_i)}$). 
A detailed description of notation is in Table~\ref{tab:notation}. 
The user can send a policy management request to the \texttt{SPM} using an input device (i.e., keyboard).
When a user requests the creation, modification or deletion of a policy, the \texttt{SPM} safely transmits the policy entered by the user to the SSD.
 
The core role of the \texttt{SPM} is to securely deliver policy information set by the user to the SSD to prevent rootkit-level malware.
By utilizing existing research such as Intel PAVP and Intel SGX~\cite{aurora, pavp}, a graphical interface safely isolated from rootkit-level malware can be provided to the user.
Also, by utilizing the existing research, Aurora~\cite{aurora} and Intel SGX, the user's input through an input device such as a keyboard can be safely transmitted to the SGX enclave.
Therefore, \sgxssd{} adopts the design of Aurora~\cite{aurora} and Intel PAVP to get a trusted screen and safely transmit user input to the enclave.

 \begin{table}[!b]
 	\vspace{-0.15in}
	\centering	
	\small
	 \caption{\small The notation defined in \sgxssd{}}
 	\vspace{-0.2in}    	
	\begin{center}
		\resizebox{\columnwidth}{!}{
			\begin{tabular}{|l||p{7cm}|c|c|c|c|c|}
				\hline
				\textbf{Notation}&\textbf{Description}\\
				\hline
				$K_{dev}$&Private key shared by \texttt{SPM} and SSD.\\
				\hline
			    $Id_{(P_i)}$&Id assigned to the policy ($P_i$) of \sgxssd{}.\\
			    \hline
			    $Id_{(f_j)}$&Id assigned to the versioned file ($f_j$) of \sgxssd{}.\\
				\hline
			    $CP_{(P_{i})}$&Configuration parameters that make up the policy($P_{i}$). It consists of Retention Time (\texttt{RT}), and Backup Cycle (\texttt{BC}).\\
				\hline
				$Rule_{(P_{i})}$&
				Policy-to-file mapping rules.\\
				\hline
			    $pMeta$&Policy hint sent by \texttt{Piggyback Module}. It consists of the policy ID ($Id_{(P_i)}$), file ID ($Id_{(f_j)}$), and file offset. \\
			    \hline
		\end{tabular}}
	    \label{tab:notation}
	\end{center}
\end{table}

The next step is to ensure that user input stored in the enclave is safely transmitted to the SSD.
For this, a secure session between the enclave and SSD is established.
The \sgxssd{} vendor inserts a symmetric key device key ($K_{dev}$) into a hard-coded form in the enclave of the \texttt{SPM} and SSD in advance. Using this symmetric key, the SSD guarantees the integrity of user input values transmitted by the enclave.
 
\subsection{Secure Delivery of Policy Hints to SSD}
SSDs have no file semantics at all.
Therefore, in order to perform per-file versioning inside the SSD, the host must provide information on the file versioning policy corresponding to each page to be updated to the SSD.
At this time, the policy metadata ($pMeta$) that the host must pass to the SSD are as follows:
(i) Policy ID($id_{(P_i)}$),
(ii) File ID($id_{(f_j)}$), and (iii) File offset.
The SSD may internally refer to (i) to find out the versioning policy corresponding to the page to be updated.
On the other hand, (ii) and (iii) are information required by the SSD to restore the file to \texttt{OV} when the file is corrupted.
For example, even though the SSD safely holds the \texttt{OV} pages of a particular file that needs to be restored, if there is no information about which offset of each \texttt{OV} page is mapped to which file, file recovery is impossible.

Two modules are implemented in the secure policy delivery framework to transfer policy metadata ($pMeta$) to the SSD: (i) \texttt{Policy Manager} and (ii) \texttt{Piggyback Module}.
\texttt{Policy Manager} is a kernel module running in the background of the server and searches policy metadata ($pMeta$) corresponding to the file.
\texttt{Piggyback Module} is an OS kernel module that transmits policy metadata ($pMeta$) to the SSD.
The \texttt{Piggyback Module} is implemented at the device driver level and system calls related to file I/O. When the application updates the file, this request must reach the SSD via the \texttt{Piggyback Module}.

\begin{figure}[!t]
	\begin{center}
		\begin{tabular}{@{}c@{}c@{}c@{}c@{}}
			\includegraphics[width=0.4\textwidth]{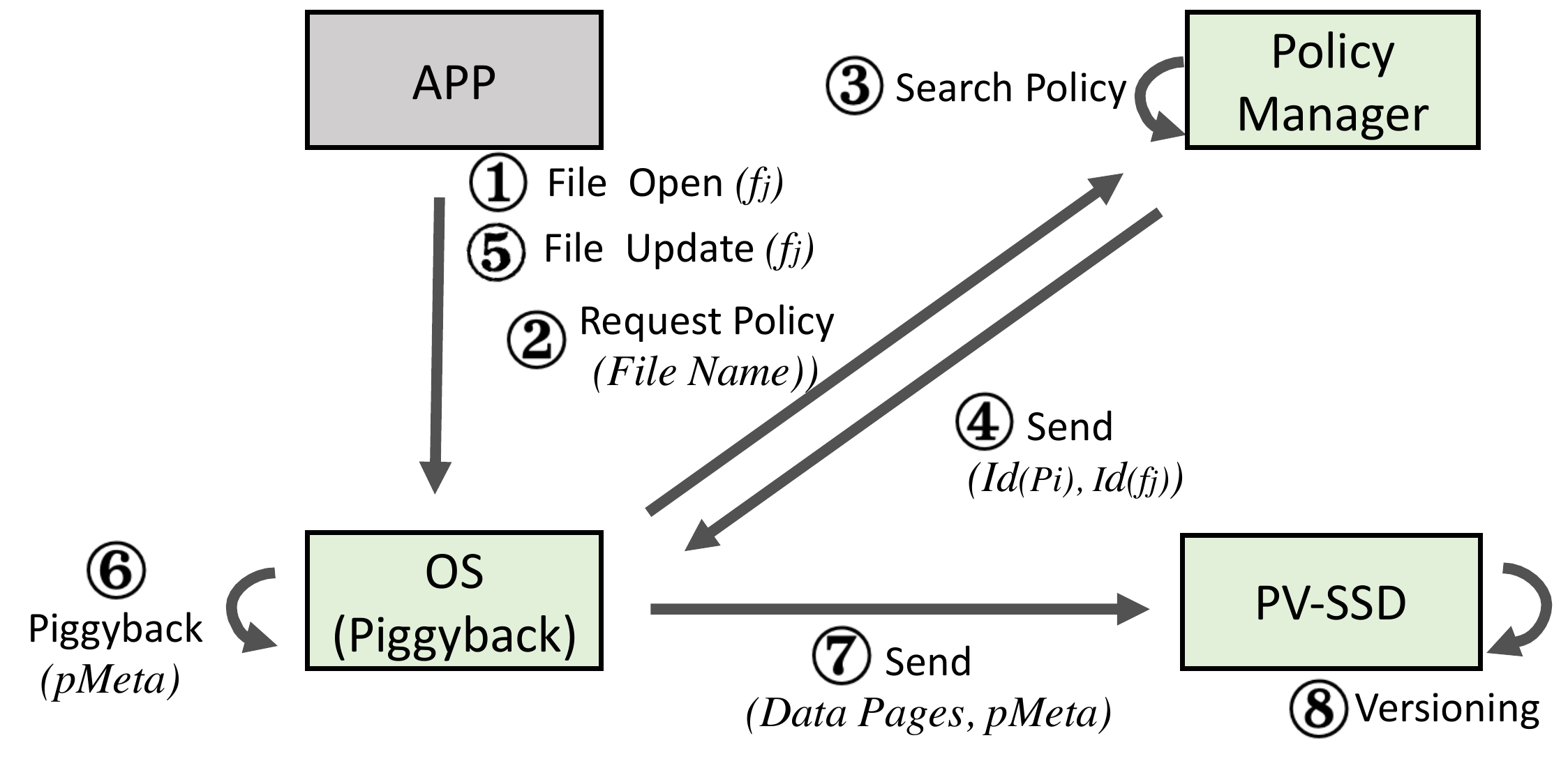} 
		\end{tabular}
 		\vspace{-0.1in}
		\caption{\small Workflow for user conf. registration and file versioning.
			}	\label{fig:workflow_versioning}
 		\vspace{-0.05in}
	\end{center}
\end{figure}

Figure~\ref{fig:workflow_versioning} shows the workflow between \texttt{Policy Manager} and \texttt{Piggyback Module}. 
First, \texttt{Policy Manager} fetches the following policy metadata from the SSD when it is first run; $Id_{(P_i)}$, $Rule_{(P_i)}$, registered list of files.
When an application opens a file ($f_j$) (\ding{172}), \texttt{Piggyback Module} sends the file name to \texttt{Policy Manager} (\ding{173}).
\texttt{Policy Manager} retrieves the policy metadata mapped to the file name (\ding{174}).
If the policy doesn't exist, a NULL value is returned to the \texttt{Piggyback Module}.
On the other hand, if the policy exists, the corresponding policy ID ($Id_{(P_i)}$) and file ID ($Id_{(f_j)}$) are sent to \texttt{Piggyback Module} (\ding{175}).

Afterward, when the application requests a file update, the update data is not versioned if \texttt{Piggyback Module} has received a NULL value from the \texttt{Policy Manager}.
However, if \texttt{Piggyback Module} has got the valid $Id_{(P_i)}$, \texttt{Piggyback Module} sends the policy information to the SSD.
When the application's file update request passes the OS(\ding{176}), it is translated to multiple per-page write requests and sent to the disk.
When the file update request is translated to multiple page write requests in the device driver, \texttt{Piggyback Module} piggybacks the policy and file information ($pMeta$) on each page (\ding{177}).
Then $pMeta$ is sent to the SSD (\ding{178}).
The performs versioning by using the received $pMeta$ (\ding{179}).

\begin{figure}[!t]
		
	\begin{center}
		\begin{tabular}{@{}c@{}c@{}c@{}c@{}}
			\includegraphics[width=0.42\textwidth]{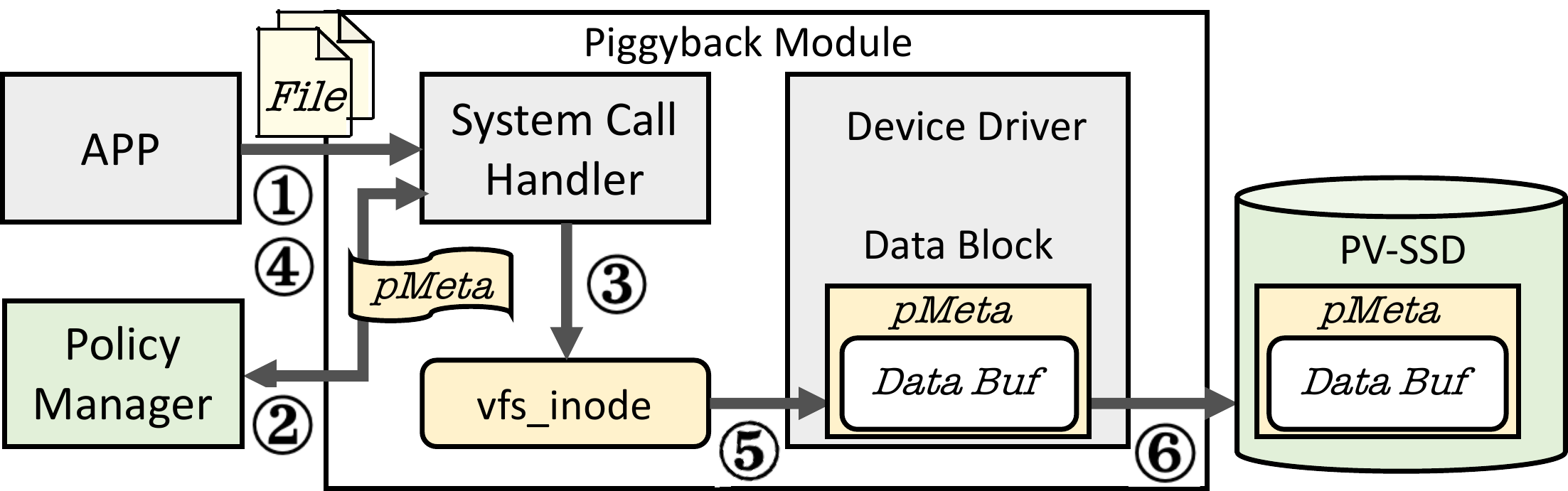} 
		\end{tabular}
		\caption{\small Piggyback Module of \sgxssd{}.
			}	\label{fig:piggyback}
	\end{center}
\end{figure}

Figure~\ref{fig:piggyback} shows the specific flow of \texttt{Piggyback Module} that sends the related policy and file information ($pMeta$) whenever an application updates the file.
First, when an application opens a file (\ding{172}), the open system call handler gets $Id_{(P_i)}$ and $Id_{(f_j)}$ from the \texttt{Policy Manager}(\ding{173}). Those values are temporarily stored in the \texttt{vfs\_inode} data structure (\ding{174}).
Those values are erased by a close system call handler when the application closes the file.
Afterward, when the application requests file update (\ding{175}), the content information of a file to be updated is translated into multiple data blocks while passing the file system and block layer, and sent to the device driver.
Before sending the data block to the SSD, the device driver gets $Id_{(P_i)}$ and $Id_{(f_j)}$ from the \texttt{vfs\_inode} data structure (\ding{176}).
Also, the file offset information is taken from the page cache that saves the file content.
The device driver piggybacks these values ($pMeta$) to the data block.
After that, it sends a block write request to the SSD (\ding{177}).
When the SSD gets the data, it refers $pMeta$ for versioning through untunneling, and writes only the data contents to the NAND flash.

\section{Selective File Versioning of SSD}

\label{sec:pvssd}
The \texttt{Versioning Module} of \sgxssd{} is implemented similarly to the block-level versioning method of the state-of-art of versioning SSD, Project Almanac~\cite{almanac}.
The main components borrowed from Project Almanac to implement the \texttt{Versioning Module} of \sgxssd{} are as follows.
(i) Time-travel Index: When a page is updated, 
it stores the previous version's page address (back pointer) in the out-of-band (OOB) area of the page~\cite{oob1, OOB2}. 
Through this process, a chain that sequentially connects all old versions 
corresponding to the logical page address (LPA) is maintained.
(ii) Compression engine: In the GC phase,  
it compresses the page to be versioned and copies it to the delta area.
The delta area is an immutable space that preserves the \texttt{OV} of a page, and is not reclaimed until the RT of all pages in a block is expired.
Using the above two components, the SSD can effectively not only store previous data versions, but also  quickly track previous pages for recovery through the chain.

\subsection{Per-File Versioning Module in the SSD}
Apart from the two components mentioned above, 
\sgxssd{} additionally implements the following three modules for per-file versioning in the SSD: 
(i) \texttt{Policy Management}, (ii) \texttt{Version Mapping}, and (iii) \texttt{PV Algorithm}.

\subsubsection{\texttt{Policy Management}}
\texttt{Policy management} is a module for managing policy metadata set by the user through the \texttt{SPM}.
First, by using the $K_{dev}$, \texttt{Policy Management} authenticates whether the request from the host actually came from the right \texttt{SPM}.
If authentication fails, the request is denied and an error message is returned.
On the contrary, if authentication succeeds, policy metadata is newly created, modified or deleted according to the user's request.

\subsubsection{
\texttt{Version Mapping}}
\texttt{Version Mapping} parses $pMeta$ sent by the \texttt{Piggyback Module}, maps to each page.
The goal of \texttt{Version Mapping} is to pre-record policy and file information corresponding to the physical page being written when the host writes the page (I/O phase).
This allows the \texttt{PV Algorithm} module to refer to the version of the page later in the GC phase to determine whether to reclaim it.
\texttt{Version Mapping} utilizes the OOB area of each page to record the file version information. 
Since OOB is an extra area allocated for each page in the SSD, 
version information of a file can be stored without additional space overhead.

If the host updates the file, \texttt{Version Mapping} runs the following flow.
When the SSD receives a page write request from the host, \texttt{Version Mapping} checks whether the $pMeta$ exists in the data page.
If $pMeta$ exists, the page should be versioned according to policy.
If this page is later overwritten, the state of this page will be in a \texttt{OV} state rather than an invalid state.
At that point, $pMeta$ is parsed and $Id_{(P_i)}$, $Id_{(f_j)}$, and file offset value are acquired.
Then, the following file version information is saved in the updated page's OOB area: logical page address ($LPA$), the time when the page is written ($WT$), the back pointer ($BP$) which is the address of previous version, $Id_{(P_i)}$, $Id_{(f_j)}$, and file offset.
However, if the $pMeta$ in the host written data page is NULL, it is the normal file write request without any configured policy.
For that, only the LPA, and $BP$ are saved in the OOB, and the page is written.

\subsubsection{\texttt{PV Algorithm}}
\label{sec:pvalgorithm}

The \texttt{PV Algorithm} is an algorithm that determines whether or not to erase the versioned pages, and runs at the GC phase.
Specifically, the \texttt{PV Algorithm} retrieves the $pMeta$ of the page recorded by \texttt{Version Mapping}, and determines whether each page should be versioned or erased through the policy information managed by \texttt{Policy Management}.

\begin{figure}[!t]
		
	\begin{center}
		\begin{tabular}{@{}c@{}c@{}c@{}c@{}}
			\includegraphics[width=0.4\textwidth]{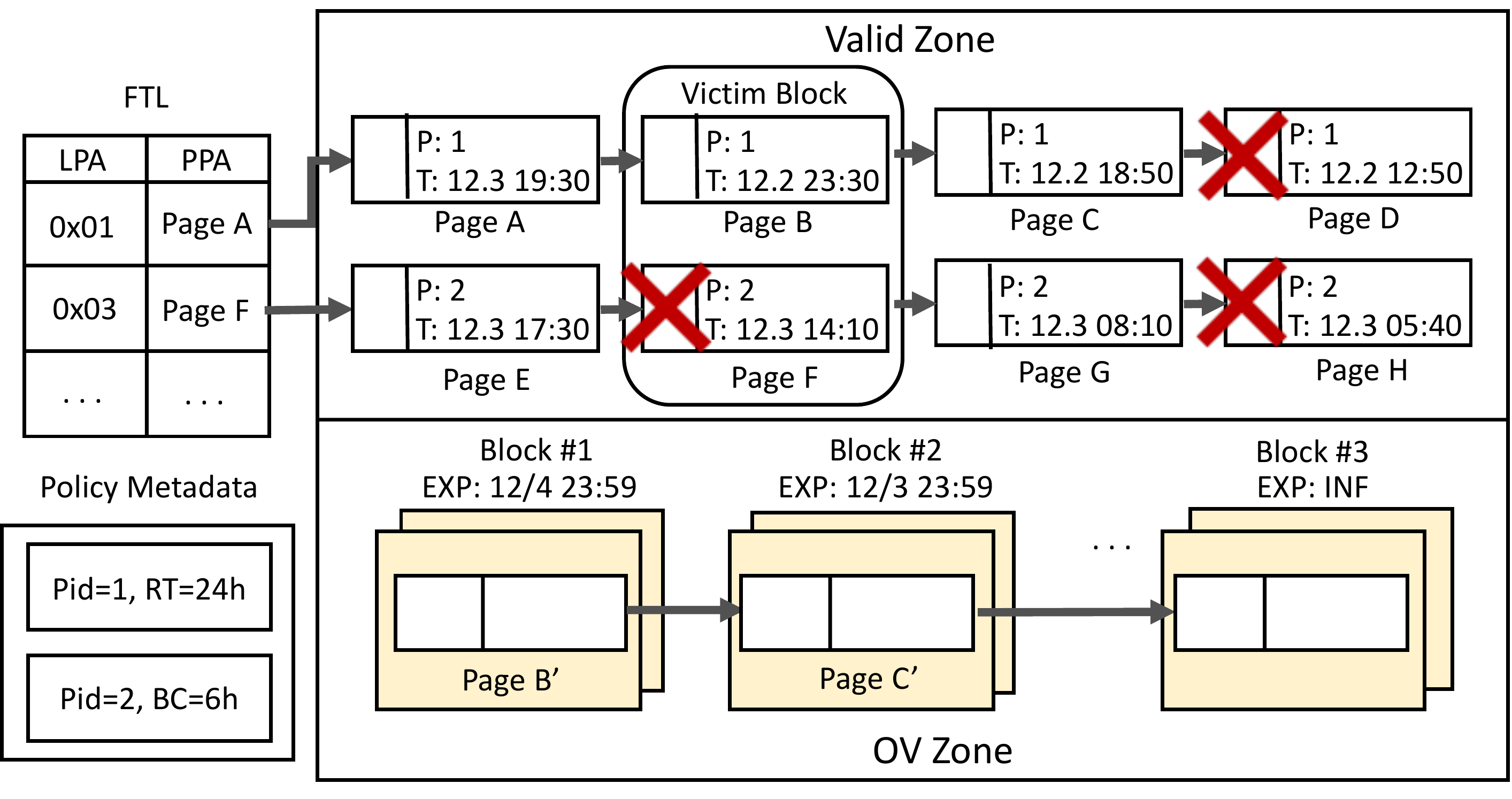} 
		\end{tabular}
 		\vspace{-0.05in}
		\caption{\small \texttt{PV-algorithm} of \sgxssd{}.
			}	\label{fig:sv}
 		\vspace{-0.25in}
	\end{center}
\end{figure}

To erase an invalidated block, the SSD performs the garbage collection~\cite{SSD}.
To erase a block during garbage collection, each page is searched to select the victim block that has the lowest rate of $valid$ pages among the saved blocks.
In a victim block, When a page is in $Valid$ state, it is copied to the free block as normal SSDs do, but the page in an $Invalid$ state is ignored.
On the other hand, when the page is in the \texttt{OV} state, the \texttt{PV Algorithm} is called. The \texttt{PV Algorithm} decides whether or not to perform versioning on the page. When searching the page in the block is done, that block is erased and free space is made.

The \texttt{PV Algorithm} performs in the following three steps:

\squishlist
\item 
{\bf Step1: File chain search:}
Figure~\ref{fig:sv} shows an image of performing versioning in the $Valid zone$ and $OV Zone$ of the SSD. 
If the victim block's page B is \texttt{OV} at the GC, LPA (0x01) is read from the OOB. After that, the latest version of the page (page A), which is mapped to FTL, is searched.
Using the back pointer stored in the OOB, all the pages (page A, B, C, and D) linked to the $Valid Zone$ are sequentially searched by traversing the version chain that begins from the page A.
While reading each page, policy metadata is searched based on the $Id_{(P_i)}$ that is stored in the OOB. 
The decision algorithm runs to determine whether to preserve each page, according to the retention period (\texttt{RT}) and backup cycle (\texttt{BC}) that are recorded in the policy metadata.
\item 
{\bf Step2: Decide whether to preserve pages:}
To determine whether to preserve each page linked to the file chain, \sgxssd implements a page preservation algorithm according the retention time or backup cycle. The page preservation algorithm is explained in detail in Section~\ref{sec:page_preserve}.

\item 
{\bf Step3: Compress and copy the \texttt{OV} pages to $OV Zone$:}
Only those pages that need to be preserved according to the \texttt{PV-algorithm} are copied to $OV Zone$.
When a page is copied to the $OV Zone$, the page is copied to the matching block according to the page's expiration time.
For example, when Page B is determined to preserved, since $Exp_{(Page)}$ of Page B is 12/4 19:30, it is copied to Block \#1.
On the other hand, since $Exp_{(Page)}$ of Page C is 12/3 23:30, it is copied to the Block \#2.
\squishend

\subsection{Page Preservation Algorithm}
\label{sec:page_preserve}
We introduce the following two page preservation algorithms: 

\textbf{{Retention Time-based Decision Algorithm:}}
Each page ($Page$) should be retained for \texttt{RT} from the time of invalidation.
The time point when the page is invalidated means the time point ($WT_{(RecentPage)}$) when the latest version of the page ($RecentPage$) is written.
Here, the latest version of the page ($RecentPage$) means the page that points ahead of the page in the file chain. 
Thus, the expiration time ($Exp_{(Page)}$) of each page can be calculated if the policy of the page is $P_i$. 

\vspace{-0.15in}
\begin{equation}
\label{eq:rt1}
Exp_{(Page)} = WT_{(RecentPage)} + RT_{(P_i)}
\end{equation}
\vspace{-0.15in}

The page should be retained if the Current Time ($CurTime$) of the SSD is smaller than the expiration time of the page. Therfore, the page is retained in the following conditions.

\vspace{-0.15in}
\begin{equation}
\label{eq:rt2}
Exp_{(Page)} > CurTime
\end{equation}
\vspace{-0.15in}

In Figure~\ref{fig:sv}, $Id_{(P_i)}$ of page B is 1. The policy metadata whose $Id_{(P_i)}$ value corresponds to 1 has an \texttt{RT} set to 24 hours.
If the current time ($CurTime$) of SSD is 12/3 20:00, Page B and C's $Exp{(Page)}$ hasn't passed the $CurTime$.
However, since $Exp{(Page)}$ of Page D is 12/3 18:50, 
it expires before the $CurTime$ comes.
Thus, Page D is erased.

\textbf{{Backup Cycle-based Decision Algorithm:}}
The SSD needs to perform versioning on data for every backup cycle (\texttt{BC}) that the user configured.
We define the backup time ($BackupTime$) as the time point when the data should be preserved.
For example, if the backup cycle is set to 6 hours, from the user's viewpoint, 06:00, 12:00, 18:00, and 24:00 daily is the $BackupTime$.
The \texttt{PV-Algorithm} decides whether the page existed in the $BackupTime$, and if it did, it performs the versioning to enable recovery to the $BackupTime$.
The generation time point ($Gen_{(Page)}$) of each page is the time ($WT_{(Page)}$) when that page was written.
On the other hand, the time at which each page has expired ($Dead_{(Page)}$) is the time ($WT_{(RecentPage)}$) when the latest version ($RecentPage$) is written.
If the backup time ($BackupTime$) is included in the page's life cycle ($[Gen_{(Page)}, Dead_{(Page)}]$), that page is preserved.
If not, that page is erased.

In Figure~\ref{fig:sv}, $Id_{(P_i)}$ of Page F is 2, and it corresponds to the policy that has 6 hours of BC. Since the lifetime of Page F is [14:10, 17:30], it is not included in the backup time and erased. On the other hand, the lifetime of Page G is [08:10, 14:10], it is included in the $BackupTime$, 12:00. Thus, it will be preserved.

If any of the conditions of the two decision algorithms described above are satisfied, the page must be preserved. On the other hand, if all conditions are not met, the page is erased.

\subsection{Recovery}

\sgxssd{} supports the file-level restoration. 
The user can use \sgxssd{} and restore files to a certain time point or version. The restoration flow is as follows. 
First, the \texttt{SPM} reads policy metadata from the SSD and shows a list of versioned files to users.
When the user decides which files to restore, the request is transferred to the SSD.
The restoration utility runs in \texttt{Fast Recovery} and \texttt{Robust Recovery} mode.
\texttt{Fast Recovery} supports fast restoration and can run in the state when the file system is not corrupted.
The \texttt{policy delivery framework} reads the LBA list of a file from the file system. Then it sends a restoration request ($Id_{(F_j)}$, restoration time point, LBA list) to the SSD. 
The SSD sequentially searches the page chain that is linked to each LBA in the LBA list that the user sent. While it is searching the chain, if the Written Time ($WT$) of a page logged at OOB are consistent with the time point that the user requested, the page and the corresponding offset information are sent to the \texttt{policy delivery framework}. 
This page is a part of the version of the file to be restored.
When the search process ends, the \texttt{policy delivery framework} receives data pages and the offset of every file version to be restored. 
The restoration process ends when the \texttt{policy delivery framework} writes these new pages with the correct offset in the recovery file.

On the other hand, if the file system is corrupted, or the file's LBA list that the file system saves has changed due to the file size being reduced by malware, file restoration is not possible with \texttt{Fast Recovery}.
In this case, \texttt{Robust Recovery} starts.
\texttt{policy delivery framework} immediately sends a restoration request ($Id_{(F_j)}$, restoration time point) to the SSD.
Then the SSD searches all the physical pages of NAND flash, and reads the OOB.
If the file ID registered in the OOB matches the file ID requested by the host, the file is restored by sending the page to be restored and the offset to the user, like the \texttt{Fast Recovery} mechanism.
This method has the disadvantage that it takes a long time to recover, but file restoration is possible even in an extreme scenario such as when a file system is compromised.

\section{Performance and Security Evaluation}

\subsection{Evaluation Setup} 

We prototyped \sgxssd{} and performed experiments in the Linux environment, based on the Intel(R) Core(TM) i7-8700 CPU @ 3.70GHz with 16 GB RAM (128 MB for EPC) that supports Intel SGX.
The \texttt{SPM} is implemented in the Enclave environment that Intel SGX supports, and the \texttt{SSD Versioning Module} is implemented in the SATA2.0-based Jasmine OpenSSD~\cite{jasmine} that has a ARM7TDMI-S with a  87.5MHz clock rate, 96 KB SRAM, 64 MB DRAM, and the 64 GB NAND Flash memory chip. 

\textbf{Implementation:} Each module was implemented as the follows. 
First, we implemented an API for two-way communication between the \texttt{SPM} and SSD.
Each module hard-coded a private key($K_{dev}$) in advance to authenticate mutual transmission.
However, for Aurora, we didn't implemented it, but rather emulated it by giving a delay.
{Also, instead of implementing a reliable display using Intel PAVP, we emulated a trusted screen.}
The \texttt{Piggyback Module} was implemented by modifying the open and close system calls and the SATA device driver layer.
The open and close system calls receive policy information by communicating with the \texttt{Policy Manager}. Furthermore, to send $pMeta$ to the SSD, the SATA device driver allocates a new page and saves the $pMeta$. 
Then the page where $pMeta$ is stored is inserted into the page list ($ahci\_sg$) to be transmitted by the DMA engine.
The DMA engine refers to $ahci\_sg$ and sends $pMeta$ along with the data pages to the SSD.

The \texttt{SSD Version Module} is implemented in the SSD firmware.
In \sgxssd{}, the implementation of version chain from $ValidZone$ to $OV Zone$ follows the design of Project Almanac~\cite{almanac}. 
For selective file version management based on policy, we implemented \texttt{Policy Management}, \texttt{Version Mapping}, and \texttt{PV-Algorithm} modules.

\subsection{Performance Evaluation}

In this section, we first analyze the performance overhead of the \texttt{SSD Versioning Module}.
Secondly, we analyze the piggyback performance of the \texttt{Policy Manager} and \texttt{Piggyback Module} in the OS kernel.
Thirdly, in a realistic usage scenario, we analyze and compare the overhead of various files versioning with other software-based versioning systems.
Lastly, we analyze the overhead that stems from the policy management of the \texttt{SPM}.
We implemented an in-house multi-process-based I/O benchmark to evaluate the performance overhead.

\subsubsection{Overhead Analysis of SSD Versioning Module}

In order to induce GC in the experiment, we created a 1GB partition on the SSD and initialized 
all pages to \texttt{Invalid Page} before each experiment.
We measured the overwriting performance of synthetic big (20 MB) and small (32 KB) file workloads (referred as $Workload(B)$, and $Workload(S)$ respectively) for performance comparison according to GC overhead.
In addition, each file has 3 days of \texttt{RT}. 
Experiments were performed while varying 
the capacity ratio and versioning ratio of the SSD.
The capacity ratio refers to the ratio of unique data, compared to the size of disk partition. 
The versioning ratio refers to the ratio of important data to be versioned, compared to total data amount.
When a versioning ratio is 1, all data will be versioned. So, it shows similar performance to full disk versioning~\cite{bvssd, almanac}.

\begin{figure}[!t]
		
	\begin{center}
		\begin{tabular}{@{}c@{}c@{}c@{}c@{}}
			\includegraphics[width=0.24\textwidth]{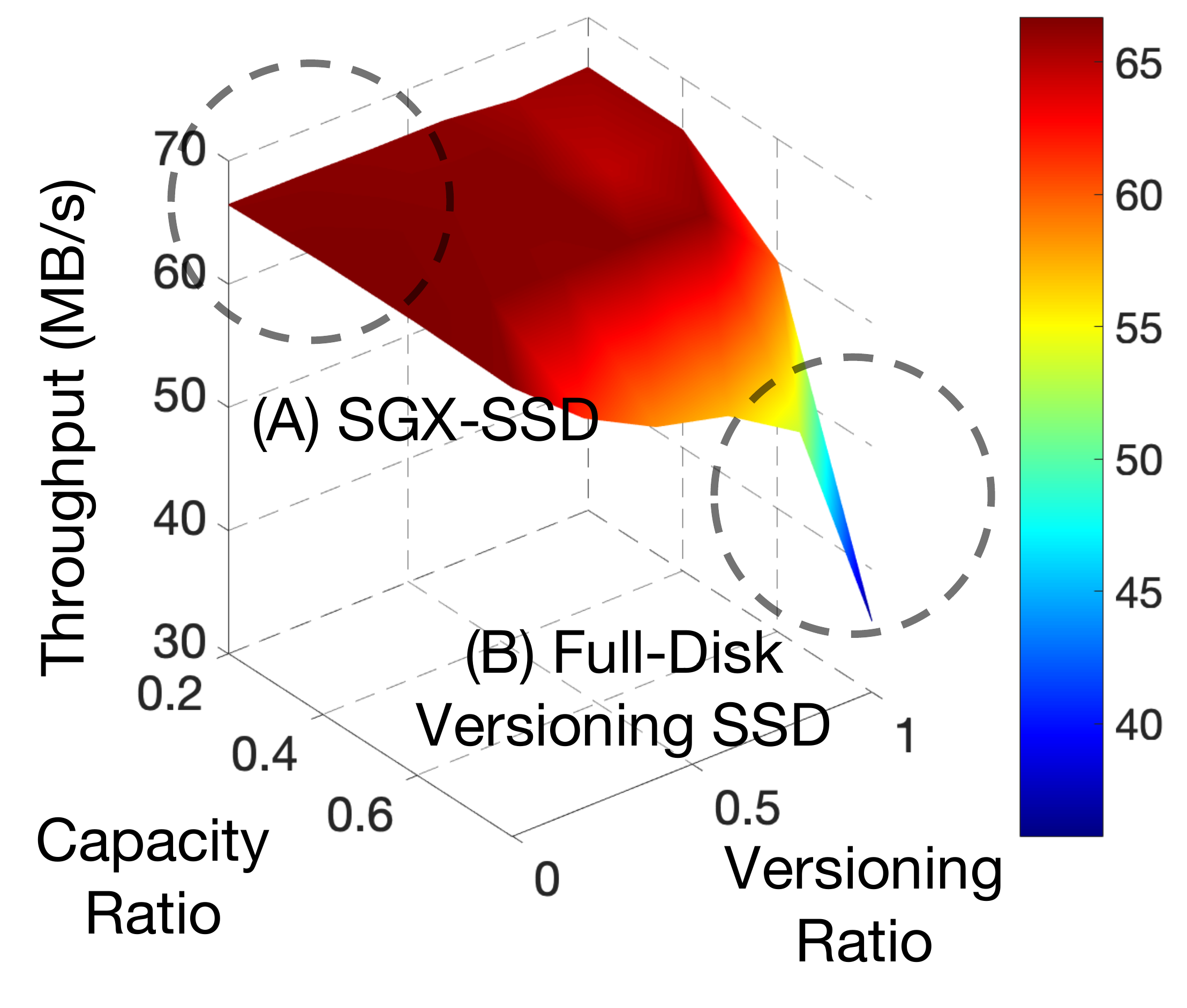} 
			&\includegraphics[width=0.24\textwidth]{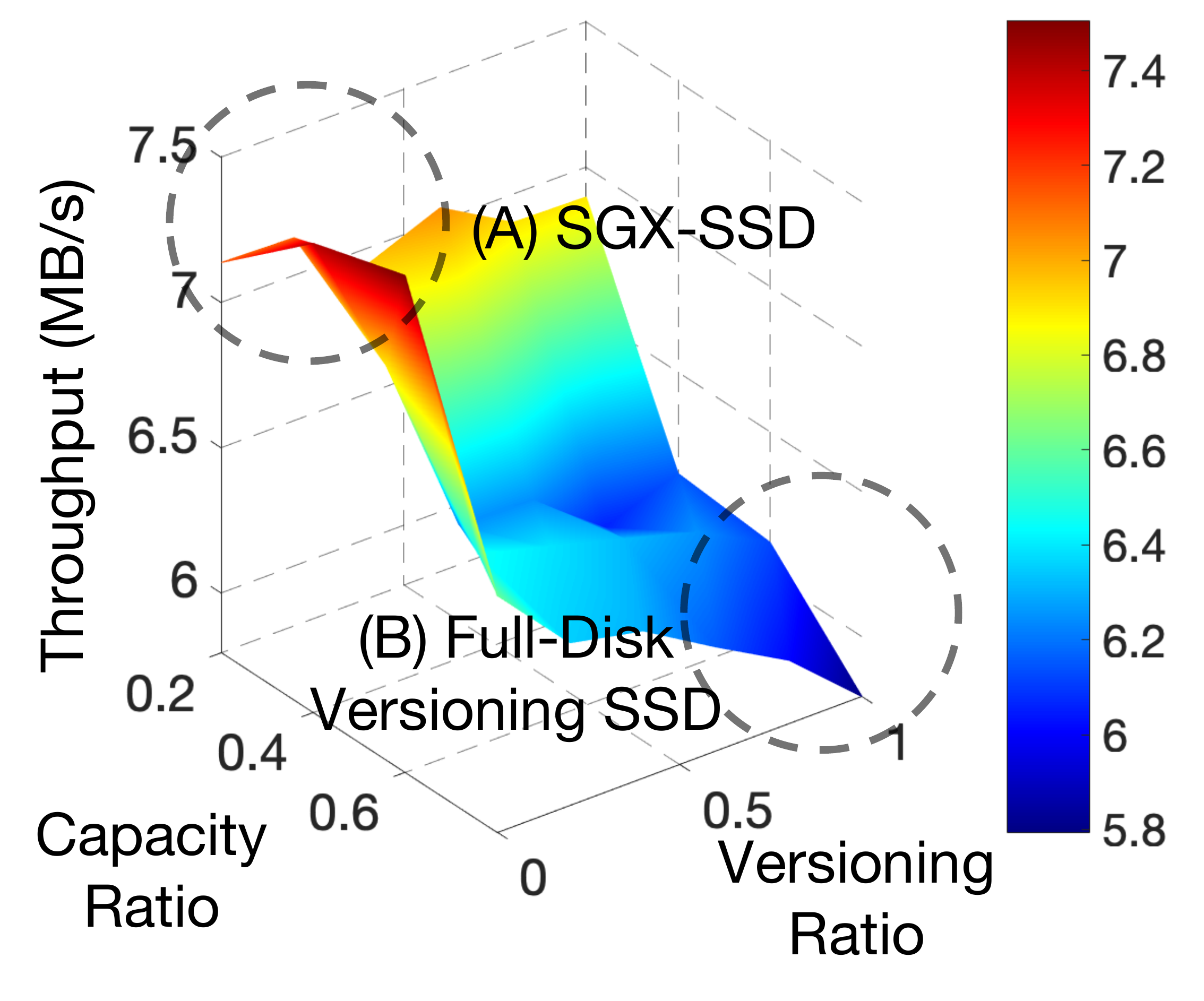}\\	
			{ \small (a) $Workload(B)$}
			& { \small (b) $Workload(S)$}\\
		\end{tabular}
		\caption{\small Performance comparison of {\sgxssd} and Full-disk Versioning SSD.
		The capacity ratio and versioning ratio refer to the unique data ratio of the total disk capacity and the ratio of versioned data among the total data, respectively. 
			}	\label{eval:perf}
	\end{center}
\end{figure}

Figure~\ref{eval:perf} shows the performance results of {\sgxssd} according to the varying capacity and versioning ratio.
In the figure, 
the dotted circles (marked as `(A)' and `(B)') indicate {\sgxssd} and full disk versioning (such as Project Almanac, respectively.
In Figure~\ref{eval:perf}(a)(b), we observe that {\sgxssd} maintains higher performance than full disk versioning SSD.
full disk versioning SSD performs versioning for all data because the \texttt{RT} of all data is the same.
On the other hand, {\sgxssd} selectively performs file versioning.
That is, with {\sgxssd}, versioning only important data is performed, so write amplification is much less than full disk versioning SSD.
Specifically, in Figure~\ref{eval:perf}(b), $Workload(S)$ lowers the internal GC efficiency of the SSD more than $Workload(B)$, so overall throughput is low.
This is because, in the case of $Workload(S)$, \texttt{OV Page} and \texttt{Invalid Page} are mixed and distributed in the same block, thereby increasing GC overhead.
In contrast, full disk versioning SSD rapidly decreases in performance as capacity ratio and versioning ratio increase, while {\sgxssd} decreases at moderate speed.

\subsubsection{Overhead Analysis of Piggyback Transmission}
\begin{figure}[!t]
		
	\begin{center}
		\begin{tabular}{cc}
			\includegraphics[width=0.23\textwidth]{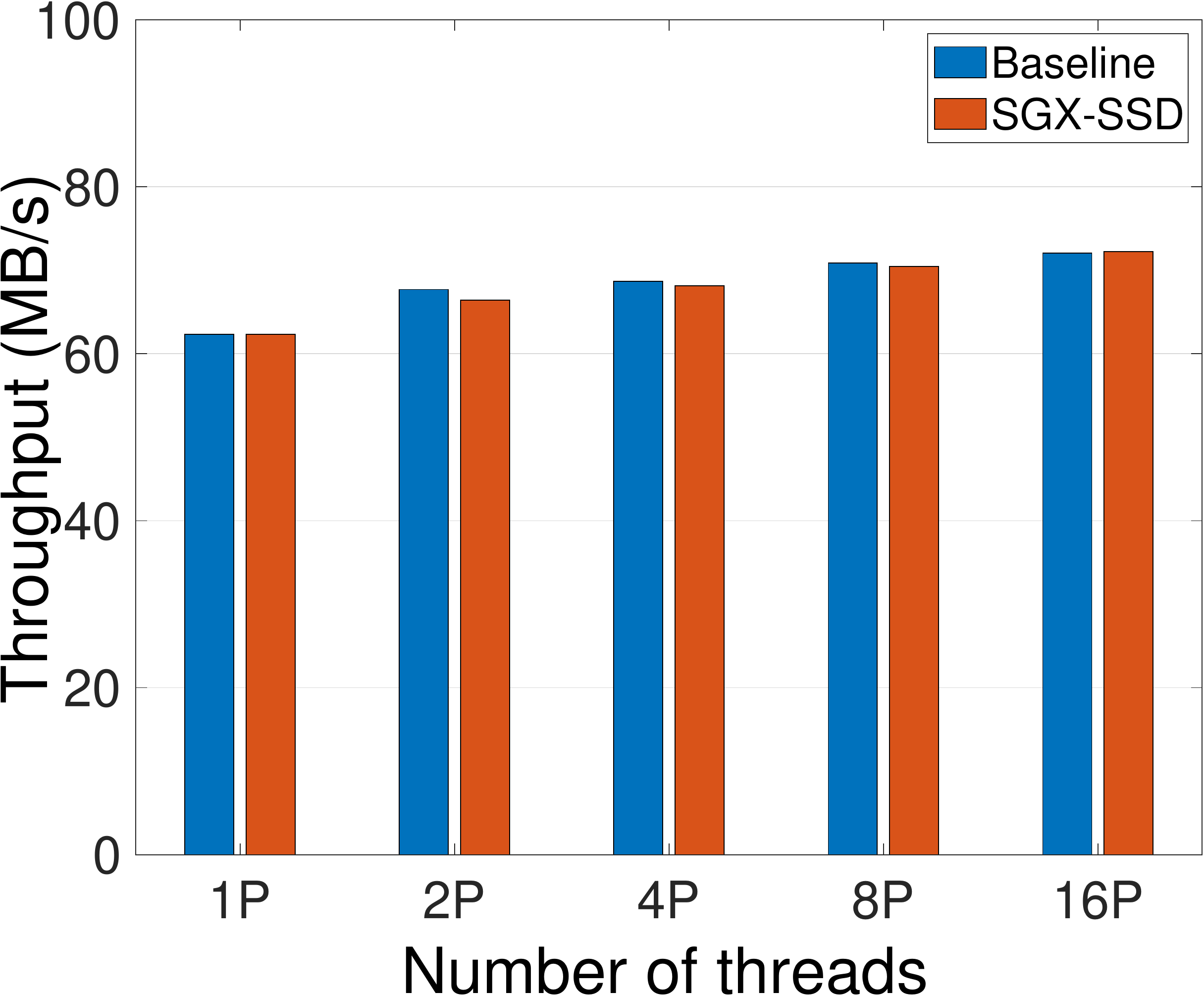} 
			&\includegraphics[width=0.23\textwidth]{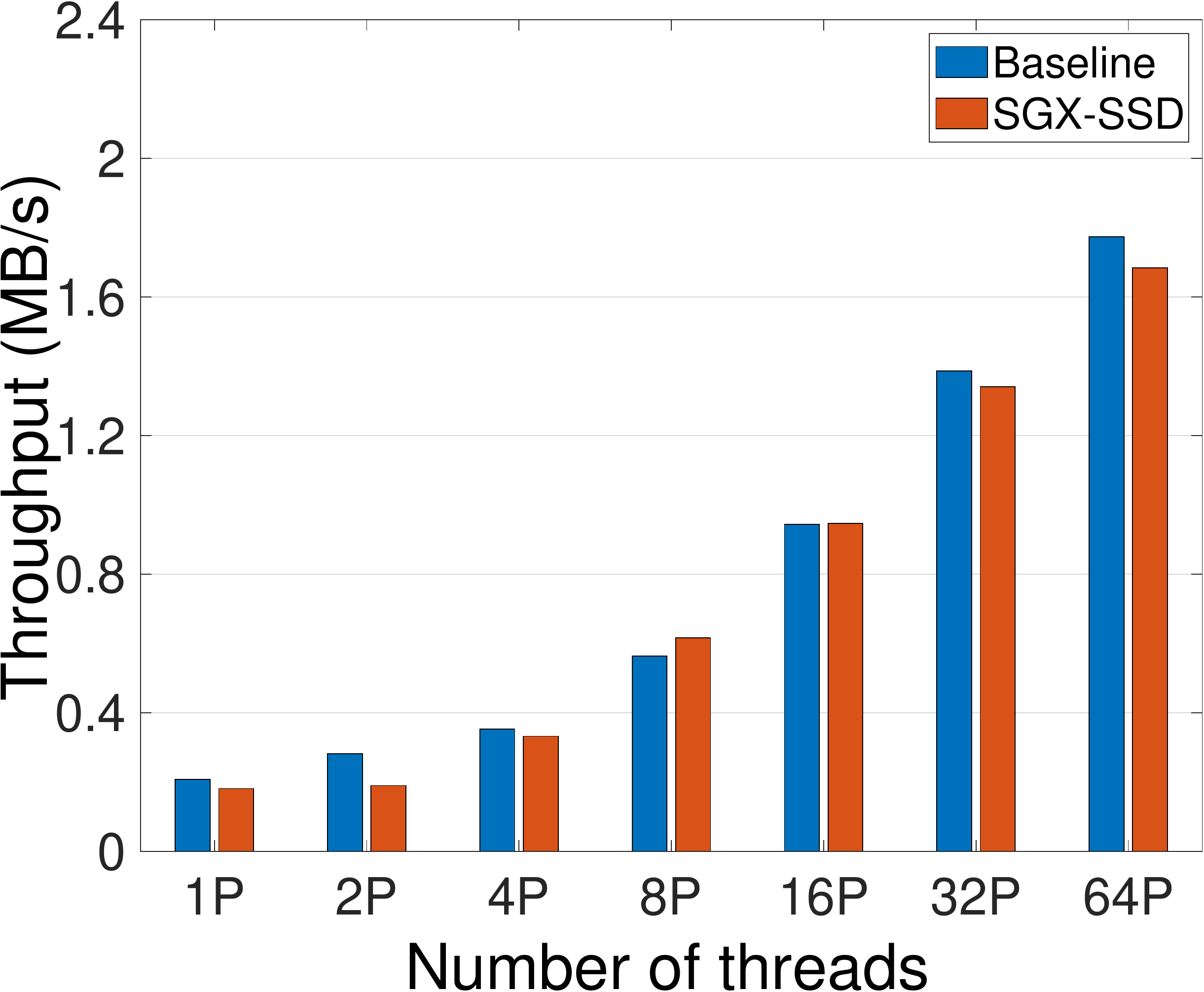}\\	
			{ \small (a) $Workload(B)$}
			& { \small (b) $Workload(S)$}\\
		\end{tabular}
		\caption{\small Perf. comparison of {\sgxssd} and Project Almanac.
			}	\label{eval:eval_piggyback}
	\end{center}
\end{figure}

In this experiment, we analyzed the overhead of the kernel implementation, which is for communicating with the \texttt{Policy Manager} and the piggyback transmission. 
Figure~\ref{eval:eval_piggyback} compares the performance results of the piggybacked I/O path (\texttt{pb\_IO}) (\sgxssd{}) and normal I/O path (\texttt{base\_IO}) (baseline).
\texttt{base\_IO} is a file write experiment that passes the normal I/O path that doesn't perform piggyback, and it is performed on the Baseline version (\texttt{Normal SSD}) of the Jasmine OpenSSD.
On the other hand, \texttt{pb\_IO} is an experiment that does versioned file write to SSD by piggybacking the policy and file information through the \texttt{Piggyback Module}, after getting the policy and file information from the \texttt{Policy Manager}.
For \texttt{pb\_IO}, a light weight interface is implemented in the Jasmine OpenSSD to get the $pMeta$ from \texttt{Piggyback Module}. 
However, to keep our focus on kernel overhead analysis, the \texttt{SSD Versioning Module} isn't implemented in the SSD for both \texttt{base\_IO} and \texttt{pb\_IO}.
That is, the SSD gets the $pMeta$, but ignores the given information and performs I/O.

Figure~\ref{eval:eval_piggyback} shows the comparison results of \texttt{base\_IO} and \texttt{pb\_IO}. Figure~\ref{eval:eval_piggyback}(a) shows the throughput when each thread writes to the 128 MB big file ($Workload(B)$). 
The evaluation result shows that the \texttt{base\_IO} and \texttt{pb\_IO} have the same performance regardless of the number of threads. This means the overhead that comes from piggybacking the policy and file information in file write is negligible. 
Figure~\ref{eval:eval_piggyback}(b) shows the throughput result when each thread writes to the 10,000 4 KB small files ($Workload(S)$). Since the file open and close overhead is included when writing to each small file, it shows the significantly lower throughput than the big file evaluation. In the evaluation result, 
most threads at \texttt{pb\_IO} have 4\% lower performance than the \texttt{base\_IO} on average.
In a file open system call, the \texttt{Piggyback Module} requests the policy and file information from the \texttt{Policy Manager} and receives them.
In this process, the overhead that comes from the \texttt{Policy Manager} searching the policy and file information seems to cause the performance drop.
However, since the rate of file open is way lower than the rate of file write in a real workload, there won't be much performance overhead considering real-world cases.

\subsubsection{Overhead Analysis of Policy Management}

To measure the overhead of the policy management, we measured the latency from the time point when the user gives a policy request as an input, to the time point when a request is sent from the Enclave to the SSD and policy modification is done.

\begin{table}[!b]
	\centering	
		\scriptsize
	\caption{{\small Latency evaluation for policy management by using the \texttt{SPM} (in ms)}.}
	\label{tab:spm}
    \vspace{-0.2in}
	\begin{center}
	    \small
			\resizebox{0.75\columnwidth}{!}{
				\begin{tabular}{|c|c|c|c|c|c|c|}
				\hline
                Policy &Creation&Change&Deletion\\
				\hline
				Metadata I/O&5.0014&5.2767&5.1860\\
				\hline
			    Data I/O&2692.4536&0&0\\
				\hline
		\end{tabular}}
	\end{center}
\end{table}

Table~\ref{tab:spm} shows the latency of a policy creation, modification, and deletion request taken to finish.
Fifty 20 MB image files (total of 1 GB) are stored on a disk in advance. 
The Metadata I/O represents the latency taken for the policy value set by the user to be reflected in the \sgxssd{} system. 
On the other hand, the Data I/O represents the file I/O latency taken to perform versioning as a configured policy.
When creating the policy, the user newly registers the policy to the SSD through the \texttt{SPM}, and registers the policy-file mapping rule ($Rule_{(P_i)}$) to the \texttt{Policy Manager} (Metadata I/O). 
Moreover, the file needs to be rewritten when versioning the pre-existing file with a created policy (Data I/O).
When the file is rewritten to the disk, since the policy and file information are piggybacked and reach the SSD, the file is versioned by the registered policy.
However, if the policy modification request is sent to the SSD through the \texttt{SPM}, the SSD only modifies the $Pmeta$. From that point, the file corresponding to the policy is managed by a modified policy.
Also, if the policy deletion request is sent to the SSD, the SSD only deletes $Pmeta$. From that point, the file corresponding to the policy is also deleted. 
Therefore, in this two cases, additional Data I/O overhead doesn't occur.
According to the evaluation results, the policy creation, modification, and deletion requests all showed about 5ms of latency. Thus, we can see that the policy management overhead is fine. 
On the contrary, the Data I/O of Policy Creation results in the 26.98 seconds of overhead while re-reading and re-writing a 1 GB file.
However, this overhead is not just bounded to the \sgxssd{}.
That's because various \textit{per-file versioning} systems~\cite{versioningFS, elephantfs} need at least one copy operation to register a version to the a normal file. 

\subsubsection{Overhead Analysis of Various Versioning System}
To prove the performance of \sgxssd{} in a realistic usage scenario, we measured the overhead that comes from 
versioning 6,517 files (total of 565.3 MB), 
in a disk's local partition, 
including the source code, document, and media files.
Figure~\ref{tab:diversesystem} shows the throughput analysis results of \sgxssd{} and various software-based versioning systems.

\begin{table}[!t]
	\centering	
	\caption{{\small Comparison of software-based versioning and \sgxssd{}}.}
	\label{tab:diversesystem}
	\vspace{-0.2in}

	\begin{center}
	    \small
			\resizebox{\columnwidth}{!}{
				\begin{tabular}{|c|c|c|c|c|c|c|}
				\hline
				&Classification&Throughput&TCB\\
				\hline
				NILFS&File system&32.82MB/s&Host server\\
				\hline
				Git&Application&24.74MB/s&Host server\\
				\hline
				Dropbox&Cloud system&6.26MB/s& Cloud servers\\
				\hline
				\multirow{2}{*}{\sgxssd}&{Device}&\multirow{2}{*}{22.64MB}/s&Device firmware\\
				&firmware &&Intel SGX\\
				\hline
		\end{tabular}}
	\end{center}
\end{table}

NILFS~\cite{nilfs} is a log-based file system 
that performs versioning on a file with a continuous snapshot. 
Dropbox is a cloud-based backup system that supports versioning. 
Lastly, Git is a locally implemented version control application~\cite{git}. 
For versioning, NILFS, \sgxssd{}, and Dropbox all copy the original files in a local partition to the secure partition that supports versioning. 
On the other hand, Git performs versioning through the add and commit commands in the local partition area.

As a result of the experiment, NILFS and Git showed about 1.4 times and 1.09 times higher performance than \sgxssd{}.
However, because NILFS and Git run at the file system and application level, respectively, if the host server is compromised, \texttt{OV}s are tampered with.
On the other hand, although Dropbox provides strong security as a cloud storage, it has the lowest throughput due to  the low network traffic bandwidth encountered while storing files on the cloud server.
\sgxssd{} has somewhat low performance when compared to the local environment software(NILFS, git). 
However, the TCB size of \sgxssd{} is much smaller than existing software-based versioning systems.
Therefore, \sgxssd{} provides strong integrity for the \texttt{OV} of files even in a compromised local system environment.

\subsubsection{Recovery evaluation}
In this evaluation, we measured the file recovery performance of {Fast Recovery} and {Robust Recovery} for a large file (20 MB) and a small file (32 KB).
We used an in-house program we developed for the recovery experiment. 
The program consists of two steps. The first step is to create a file and write to the file for the first time. 
Then, the file is overwritten with encrypted values to assume a tamper attack by malware.
For recovery, the next step is to call the file recovery API using the policy delivery framework, and measure the time it takes to recover the modified file. 
The SSD is assumed to be a 1 GB partition.

\begin{table}[!b]
	\centering	
	\caption{{\small Recovery rate and performance evaluation of \texttt{Fast Recovery} and \texttt{Robust Recovery}.}}
	\label{tab:recovery}
	\vspace{-0.1in}

	\begin{center}
			\resizebox{0.85\columnwidth}{!}{
				\begin{tabular}{|c|c|c|c|c|c|c|}
				\hline
                &Recovery rate&Big file&Small file\\
				\hline
				Fast Recovery&100\%&69.563ms&6.825ms\\
				\hline
			    Robust Recovery&100\%&13.194s&12.728s\\
				\hline
		\end{tabular}}
	\end{center}
\end{table}

Table~\ref{tab:recovery} compares the recovery time of Fast Recovery and Robust Recovery.
First of all, 
both showed a 100\% recovery success rate.
Specifically, the Fast Recovery experiment showed that the recovery time of the small file is about 10 times faster than that of the big file.
This is because the larger the file size, the greater the number of pages that the SSD has to search to restore.
On the other hand, the performance of Robust Recovery is about 190 times lower than that of Fast Recovery.
In Fast Recovery, the SSD performs restoration by selectively searching the chain only for pages mapped to the LBA list sent by the host.
On the other hand, in Robust Recovery, the performance is much lower because the SSD has to scan all physical pages of the entire SSD partition.

\subsection{Security Analysis}

For thorough security analysis, we divided the modules in \sgxssd{}, which are \texttt{SPM}, \texttt{Policy Manager}, \texttt{Piggyback Module}, and the \texttt{SSD Versioning Module} into each attack vector and analyzed the danger of malware's file corruption that can happen in each attack vector.

\subsubsection{Attack Vector on \texttt{SPM}}

\squishlist
\item
\textit{Policy Modification and Deletion:}
A user can modify and delete the policy by using the \texttt{SPM}. If the malware can modify or delete the policy, it will be the main target of malware and the danger exists that file's \texttt{OV} can be erased. 
Note that, as mentioned in the threat model in Section~\ref{sec:threat}, \sgxssd{} only guarantees the integrity of pre-existed data versions that had existed before the system was infected by the malware. To be more specific, 
the malware can impersonate an authenticated user and request the \texttt{SPM} to modify or delete the policy. Also, it can be disguised as the \texttt{SPM} and request the disk to modify or delete the policy.
However, the \texttt{SPM} suspends the OS through Aurora, and authenticates and only receives input from 
a trusted user's input device (keyboard) in SMM mode.
Thus, even rootkit malware can not deceive the \texttt{SPM}.
Also, since a two-way authentication module is implemented between the \texttt{SPM} and \texttt{SSD Versioning Module}, the request sent from \texttt{SPM} is safely authenticated in the \texttt{SSD Versioning Module}.
Therefore, requests sent from disguised \texttt{SPM} are all rejected by the \texttt{SSD Versioning Module}.

\item
\textit{Disguise as SPM to deceive users:}
Malware disguises itself as the \texttt{SPM} and can trick users and induce them to modify or delete incorrect policies.
By using the existing Intel PAVP~\cite{pavp} and SGX technology, a trusted channel can be formed between the display and Intel SGX to provide a trusted display to users.
If a secret symbol that only the user knows is inserted in the trusted display, the user can distinguish whether the \texttt{SPM} screen is trustworthy by checking the secret symbol.
\squishend

\subsubsection{Attack Vector on \texttt{Policy Manager}}
\squishlist
\item
\textit{$pMeta$ tampering attack:}
Since \texttt{Policy Manager} is a utility that runs at the user level, it is vulnerable to malware attacks. However, even if the \texttt{Policy Manager} is attacked, \sgxssd{} guarantees the integrity of file \texttt{OV}s. 
Malware can forge a policy-file mapping rule ($Rule{(P_{i})}$) information on \texttt{Policy Manager}. Also, it can deceive the user and disguise itself as the \texttt{Policy Manager}, and send the wrong policy information to the \texttt{Piggyback Module}.
In these two cases, since the \texttt{Piggyback Module} receives the wrong policy information ($pMeta$) from the \texttt{Policy Manager} and sends it to the \texttt{SSD Versioning Module}, the newly updated file information is not properly versioned.
Nevertheless, the pre-existing file version that was already updated to the disk before the malware invasion is safe from corruption. Thus, restoring the data to the time point before malware breaks in is possible.

\item
\textit{Manipulating the list of files: }
When a new file mapped to a policy is created, the \texttt{Policy Manager} notifies the SSD by sending 
information about the newly added file to the SSD. 
The SSD adds new files to the list of files corresponding to that policy. 
However, 
malware can send the wrong file information to the SSD, and incorrect file information may be recorded in the file list. 
At this moment, if the users enables the \texttt{SPM}, wrong file information may be added to the file list.
However, this method cannot prevent users from recovering the \texttt{OV} of existing files.
The update request of the list of files sent by the \texttt{Policy Manager} to the SSD is limited to append only, and it can not modify or delete files registered in the list of files.
Therefore, malware cannot make a request to the SSD to change or remove the existing list of files.
\squishend

\subsubsection{Attack Vector on \texttt{Piggyback Module}} 
Since the \texttt{Piggyback Module} is a kernel module, it is vulnerable to rootkit malware. For example, malware can launch a man-in-the-middle attack at the vfs\_inode structure or the device driver layer and delete the $pMeta$ information sent by the \texttt{Piggyback Module}. In this case, the newly updated file information arrives at the \texttt{SSD Versioning Module} without any policy and is not versioned.
However, as explained above, the pre-existing file version that was updated to the disk before the malware invasion is not corrupted. Therefore, restoring the data to the time point before a malware breaks in is possible.

\subsubsection{Attack Vector on \texttt{SSD}}

Malware can corrupt or delete the files by overwriting the important files saved in the SSD. If malware overwrites a file, the original file stored in the \texttt{SSD Versioning Module} is preserved by versioning, restoration to the original data is possible.

\subsubsection{Attack Vector on \texttt{Operating System}} Rootkit malware can attack the files by compromising the OS, corrupting the file system, or corrupting the metadata that stores the LBA information of a file.
However, the \texttt{SSD Versioning Module} doesn't depend on the file system at \texttt{Robust Recovery} mode and restores files by only using the $Id_{(f_j)}$ and the file offset information stored in the OOB.

\section{Related Work}
\label{sec:related}

\textbf{Versioning Software:}
The previous backup software performs backup in (i) local storage; (ii) network-attached remote storage; and (iii) cloud storage(e.g., Dropbox, Google Cloud).
(i) and (ii) are a major attack targets of the privilege escalated ransomware~\cite{ryuk2}. 
Though (iii) can protect backup data from client-side ransomware, 
TCB is much larger than \sgxssd{} due to a huge network-connected server.
On the other hand, the trust of \sgxssd{} is limited to the Enclave that provides TEE and the SSD's firmware space, thus providing a much smaller TCB.
 
\textbf{Device-level Data Protection:}
Pesos~\cite{pesos} is access-control-based object storage and provides a richer policy.
However, the security scope of Pesos is limited to remote servers, and assumes that the client system is trusted. Therefore, data cannot be protected when malware enters the client machine.
Project Almanac\cite{almanac} versions all files at the disk level to protect data from privileged malware.
Inuksuk~\cite{inuksuk} uses Intel TXT and self-encryption disk (SED) to protect data by copying it to a protected partition.
Inuksuk, like \sgxssd{}, can selectively protect files in the disk.
However, Inuksuk has a very large overhead when backing up data to a protected partition (e.g., 23.38 seconds to back up 85.6 MB of JPG files~\cite{inuksuk}), and the whole system is interrupted while data is being backed up.
Due to the long system downtime problem, Inuksuk is unable to back up data in real time, and the size of the files to be backed up and backup cycle are limited.
In particular, in the client-server model, 
which is very sensitive to service interruption, 
an Inuksuk server must be additionally deployed to prevent service interruption.
On the other hand, \sgxssd{} not only performs host service and {\texttt{PV-SSD}}'s versioning at the same time, but also has very little overhead caused by versioning.

\section{Conclusion}

In this paper, we analyzed the security limits of the previous versioning SSD that performs full disk versioning, and propose the \sgxssd{}, a per-file Versioning SSD, to overcome the limits.
The existing full disk versioning SSD can not guarantee the integrity of data from multiple malware due to the limited retention time. 
The \sgxssd{} proposed in this paper provides the user with a policy for fine-grained versioning, and so the user can selectively keep a deep history only for the important files. 
For this, \sgxssd{} has a secure host interface to securely get the policy information from the user. 
Also, a piggyback module is designed to tell the file semantics to the SSD without additional overhead, 
and an algorithm is designed in the SSD to selectively preserve the file version based on the policy.
In addition, we conducted various security analysis evaluations and proved the high security of \sgxssd{}.

\section*{Acknowledgment}
{
This research was supported by Samsung Semiconductor research grant.
}


\bibliographystyle{IEEEtran}
\tiny
\bibliography{ref}

\end{document}